\documentclass[prd,twocolumn,tightenlines]{revtex4}
\usepackage{amsmath}
\usepackage{amssymb}
\usepackage{graphicx}
\usepackage[retainorgcmds]{IEEEtrantools}
\usepackage{bm}

\newcommand{\be}{\begin{equation}}
\newcommand{\ee}{\end{equation}}
\newcommand{\bea}{\begin{eqnarray}}
\newcommand{\eea}{\end{eqnarray}}
\newcommand{\ba}{\begin{eqnarray}}
\newcommand{\ea}{\end{eqnarray}}

\begin{document}

\title{The W-Z-Top Bags}
\author{Marcos P. Crichigno$^1$, Victor V.Flambaum$^2$, Michael Yu.Kuchiev$^2$
and Edward Shuryak$^1$}

\affiliation{$^1$Department of Physics, State University of New York,\\
Stony Brook, NY 11794, USA}
\affiliation{$^2$School of Physics, University of New South Wales, Sydney\\
2052, Australia}
\date{\today}

\begin{abstract}

We discuss a new family of  multi-quanta bound states in the Standard Model, which exist due
to the mutual Higgs-based attraction of  the heaviest members of the SM, namely, gauge quanta $W,Z$ and (anti)top quarks,  $\bar t, t $. 
We use a self-consistent mean-field approximation, up to a rather large particle number $N$. In this paper we do not focus on
weakly-bound, non-relativistic bound states, but rather on ``bags" in which the Higgs VEV is significantly modified/depleted.
The minimal number  $N$ above which such states appear strongly depends on the ratio of the Higgs mass to the masses of $W,Z,\bar{ t}, t $:
 For a  light Higgs mass  $m_H \sim 50\, GeV$ bound states start from $N\sim O(10)$, but for a
``realistic" Higgs mass, $m_H\sim 100\, GeV$, one finds metastable/bound 
 $W,Z$  bags only for $N\sim O(1000)$. We also found that in the latter case pure top bags disappear for all N, although top quarks can still be well bound to the W-bags.  Anticipating cosmological applications (discussed in a companion paper) of these bags  as  ``doorway states" for baryosynthesis , we also consider the existence of such metastable bags at  finite temperatures, when SM parameters such as Higgs, gauge and top masses are significantly modified.    

\end{abstract}

\maketitle

\section{Introduction}
\subsection{Motivation}

In the Standard Model (SM), the interaction of particles includes an attractive Higgs exchange. 
For a two-particle system it is not difficult to see under which condition a Higgs exchange would lead to bound states of such particles. Unfortunately, one finds that the corresponding critical Higgs mass lies far below the current experimental bound  $m_H^{exp}\gtrsim 116\,GeV$. 
But one should not be discouraged too early by this example. Being a scalar, the Higgs generates {\em universal attraction} between all kinds of particles. Furthermore, the strength of the attraction is proportional to their mass, similar in this respect to the gravitational force. 
Gravity, feeble as it is, is able to hold together planets, stars and even create closed systems (black holes), because the rather weak coupling can be compensated by a large
number $N$ of participating particles. Unlike vector-field based forces induced by electric, weak  or color charges, both gravity and scalar exchanges are exempt from ``charge screening"  and  become increasingly stronger for large number of particles. However, there is an important difference with gravity in that the Higgs boson is neither massless, nor particularly light in comparison to $W,Z$ or $t$. This leads to
the following question: What  happens with heavy multi-quanta states when the Higgs mass is increased, from a near-zero value, to $M_H\sim O(100\, GeV)$  where  it may be soon found. This is the main subject to be addressed in this work. 

An instructive analogy is provided by nuclear physics. 
It is convenient to describe such situation by a 
  (much-simplified) Walecka model, in which the nuclear forces can be approximately described by the $\sigma$ and $\omega$ meson exchanges.
  The  correlated two-pion state $\sigma$  is ``the Higgs boson of the nuclear physics", obtaining VEV in chiral symmetry breaking.
   The $\sigma$ and $\omega$ meson masses set a scale $1/m_{\sigma,\omega}\sim 0.3 \, fm$ 
  for the range of nuclear forces. Naively it is  too small compared to nuclear sizes (several $fm$) or  to the typical inter-nuclear distances $n^{-1/3}\sim 1.5\, fm$. Furthermore,
  because of similarity of masses $m_\sigma\sim 600\, MeV,m_\omega\sim 770\, MeV$, as well as couplings, the
  sigma-induced attraction  is nearly canceled by the omega-induced repulsion. The sum is an
   order of magnitude smaller than one would get from scalar and vector components taken separately.
  The resulting  nuclear forces thus miss to bind two neutrons and can
   barely bind a deuteron.
   And yet there are lots of bound/metastable nuclei, a subject of active branch of physics for the last century.
   
   Is the situation at electroweak scale similar? It is widely believed that the Standard Model is just a
low energy effective Lagrangian, hiding  important physics behind
its simplistic scalar Higgs. Indeed, both in superconductivity
and in QCD chiral symmetry breaking, there are no fundamental scalars,
and the condensates are in fact the two fermion or fermion-antifermion
combinations, respectively.The fermionic
mass generation by the Yukawa couplings is perhaps also just a parameterization,
soon to reveal its  dynamical content in coming LHC experiments. We are not going to speculate on this issue which has huge literature,
 mentioning only one thing. We are not aware of any particular model which would  proposes a vector companion to Higgs
with a similarly small $O(100\, GeV)$ mass.   For example,  the ``techni-$\rho$" is predicted to be at the scale
$M_\rho=1-2\, TeV$, see e.g. \cite{Kurachi:2006ej}). Thus, unlike in the nuclear physics,
 one is $not$ expecting the scalar-vector cancellation to take place. Heavier exchanges with a TeV scale mass can be included
as perturbation later, and for now effective description of the SM Higgs seems to be quite adequate for the purposes of this paper.

In this work we study the  binding of various multi-quanta states within the SM, both semi-analytically for a large number of particles $N\rightarrow\infty $, and numerically for finite $N$. We will discuss the conditions under which metastable and bound  states of $N$ heavy quanta occurs, 
considering heavy fermions (top quarks) and gauge bosons ($W,Z$). We study these objects in vacuum (zero temperature/density), as well as at finite temperature, envisioning possible cosmological applications.  The method used is the mean-field approximation, ignoring $O(1/N)$ effects such as
center of mass motion. 
Unlike previous works,  we do not focus
on weakly bound nonrelativistic systems but rather on ``bags" in which the Higgs VEV inside is strongly modified.
(One strong motivation for that is related with occurrence of the
electroweak sphalerons inside them, see the companion paper \cite{FS_baryo} requiring near-empty bags.)
We will see that bags containing W-bosons always exist, regardless of the value of the Higgs mass, provided there's a large enough number of them. Next, we study pure top bags and find that for a wide range of $N$ pure (anti-)top bags  at realistic Higgs mass are excluded, even as metastable 
minima. 

\vspace{5mm}

Why should one study these multi-quanta states?
From a methodical point of view, they are a new class of manybody systems, beyond atoms and nuclei, which are truly relativistic.  Although the experimental production of a large number of heavy
quarks/bosons in a sufficiently small volume is clearly impossible, at the LHC or any other proton
accelerator , their existence  may be important cosmologically. In fact, in our companion paper \cite{FS_baryo} we show that such bound states
may  play a significant role in cosmological Baryogenesis, provided they significantly deplete the Higgs VEV and are sufficiently long-lived. We specifically consider a scenario, studied theoretically and numerically for some time, in which the transition to the current broken phase of the SM happens directly from inflation reheating, avoiding the electroweak phase transition. 

\subsection{Recent works}

Having outlined the main issues to be addressed in this paper, let us provide an overview on recent related studies.  The existence of bags due to a Higgs-induced attraction was discussed long ago for \textit{light} quarks  in \cite{Shaposhnikov_bags}, in which case the critical number $N_c$ for these objects to exist is astronomically large. In order to reduce this number, one should look for
particles with a stronger Higgs coupling, i.e. a larger mass: thus this paper is
focused on the heaviest members of the SM, the top-quark and the W-boson.   The interest in the
issue of ``top bags" originated from the question whether a sufficiently heavy
SM-type fermion should actually exist as a bag  state,
depleting the Higgs VEV around itself 
\cite{JS}.  Although classically this seemed to be possible, it was 
shown in refs \cite{Dimopoulos:1990at, Bagger:1991pg,Farhi:2003iu} that quantum (one loop)
effects destabilize such bags, except at so large coupling at which the
Yukawa theory itself becomes apparently sick, with an instability of its ground state.
The issue rest dormant for some time till Nielsen and Froggatt \cite{Froggatt:2008ns} suggested to look at the first magic number, 12
tops+antitops  corresponding to the maximal occupancy of the lowest $
l=0,\, j=1/2$ orbital, with 3 colors and 2 from $t+\bar{t}$. Using simple
formulae from atomic physics these authors suggested that such system
 forms a deeply-bound state. 
  In ref.\cite{Kuchiev:2008fd} some of us have checked this claim and found that, unfortunately, this is $not$ the
case. While for a massless Higgs there are indeed weakly bound states of 12 tops, 
 for a realistic Higgs mass
 they disappear.  Further  variational improvement of
the binding conditions for the 12-quark system were discussed by Richard  
\cite{Richard:2008uq} who confirmed that 12 tops are unbound for Higgs mass 
$m_H>M_c(12)\sim  50\, GeV$ or so. The issue has been also discussed further in the second paper of the same authors \cite{KFS}, in the direction of a hypothetical 4th generation
of SM superheavy fermions. In
particular, a possibility that baryons ($N=3$) made of such
fermions could be lighter than the sum of their masses, and possibly produced before the fermions themselves.



  Let us also mention a number of papers
discussing more general theoretical problems in the
Standard Model. We wish to mention two of them,  specifically tied to the top quark and its relatively large Yukawa coupling. These are the issues of (i) quantum corrections and (ii) the so-called ``top condensation". 

Regarding (i), the magnitude of quantum corrections to \textit{single}-top bags has been addressed by several authors. It is known that quantum corrections can deflate single-top bags into inexistence, or even turn the vacuum unstable! (see, e.g.  \cite{Dimopoulos:1990at, Bagger:1991pg, Farhi:1998vx}).  Similarly, the role of quantum corrections may be crucial to the existence of \textit{multi}-quanta bags. This led two of us \cite{Crichigno:2009kk} to study the magnitude of one-loop quantum corrections to such bags in a scalar approximation (which ignores the spin of the top quark and the Pauli principle) as a function of $N$ and $g$. Specifically, by writing the quantum-corrected energy of such states as a loop expansion
\begin{equation}  \label{eq_Ck}
E(N)=E_{cl.} \left[ 1+ \sum_k C_k(N) ({\frac{g^2 }{4\pi}})^k \right],
\end{equation}
we calculated the one-loop coefficient $C_1(N)$,
following both the specific model and the practical tools developed in 
\cite{Farhi:2003iu}. The central issue  was the $N$%
-dependence for different couplings:  explicit calculation of $C_1(N)$ for $g=1..8$ showed that the for $g\sim1$, the correction for large bags is dominated by the scattering phases of vacuum fields on the
bag, and that $C_1\sim O(1)$ for large $N$ without any strong $N$-dependence. For top
quarks (the Yukawa coupling is  $g^2/4\pi\approx .08$) the one-loop quantum 
corrections are about $\Delta E=0.06 E_{cl.}$. We thus concluded that for $W,Z$,top-bags they are under good control,
although for hypothetical 4-th generation fermions
with a coupling a few times larger (i.e. mass $> 1\, TeV$) quantum corrections get large.

The  second issue is the  idea of top condensation  (for review and references see e.g.  \cite{Cvetic:1997eb}). A (hypothetical) strong
 attraction in the $\bar t t$ channel can lead to chiral symmetry breaking and a nonzero  $<\bar t t >$ condensate. The  lowest scalar meson would  be identified with the Higgs  boson. Whether the Higgs is composite or not is unimportant at low density, at the onset of multi-top bound states .   However, as we increase the number of quanta and get  deeper bound states
     the top density grows  and one should confront the issue of  the ultimate fate of the top bags. 
   Two logically possible scenarios are (a) saturation at a certain finite positive energy per particle, analogous to, e.g., nuclear matter, or (b) creation of a ``chirally restored vacuum" with zero chiral condensate  $<\bar t t>=0$. 
Fortunately, as we will show below, the W,top bags prefer configurations in which the density at large $N$ is saturating, thus there seems to be no need to discuss a possible chiral transition and other issues related with it.

\subsection{The main approximation and the plan of the paper}
Trying to make the paper as simple as possible, we ignore many secondary issues which can be easily included. In particular 
we ignore all forces other than Higgs-induced ones, such as e.g. strong interaction between top quarks or electromagnetic energy related to total charge of the bag. We set to zero the Weinberg angle, making $Z$ degenerate with $W$. (In fact we will call below all gauge quanta $W$, for brevity.)
As we deal with large enough particle numbers $N$, one may assume that total electric, electroweak and color  charges of the bag can be made sufficiently close to zero to ignore their mean fields.  
The lifetime of $W,Z,t$ in the bag will not be discussed in this paper, but it would be delegated to a companion paper \cite{FS_baryo}. 

The remainder of this paper is structured as follows. In section \ref{Qualitative_discussion} we give a qualitative discussion of multiple types and properties of bags. In section \ref{sec_bosons} we discuss W-boson bags, for which additional technical details are included in Appendix \ref{Appendix_bosons}. Then, in section \ref{sec_fermions}, as well as in Appendix  \ref{Appendix_fermions}, we study top quark bags and in section \ref{Mixed} bags containing both W's and tops. In section \ref{sec_T} we discuss finite temperature effects and we finally conclude with some comments about cosmological applications. 

\section{Qualitative discussion of bags}
\label{Qualitative_discussion}

\subsection{Creating a bag in the Higgs vacuum}

According to the SM, the dynamics of the Higgs field $\Phi $ is described by the Lagrangian density 
\be{
\mathcal{L}_{Higgs}=\frac{1}{2} |D _{\mu }\Phi |^{2}-\frac{\lambda}{4} \left(\Phi \Phi ^{\dagger }-v^{2}\right)^{2},
\label{Higgs_Lagrangian}
}\ee
where $D_{\mu}$ is the corresponding (gauge) covariant derivative. The usual vacuum state therefore corresponds to a homogeneous Higgs field $\left\langle \Phi \right\rangle=\bold{v}$ permeating all of space.
In the conventional unitary gauge, the Higgs field $\Phi$ is represented by the real, dimensionless, field $\phi$
\begin{equation}
\Phi\,=\,v \left( 
\begin{array}{c}
0 \\ 
\phi%
\end{array}%
\right), \label{Phi-xi}
\end{equation}
where $v$ is the vacuum expectation value (VEV) which the Higgs achieves when $%
\phi=1$. Assuming spherical symmetry, the energy reads

\begin{equation}
E_{Higgs}=2 \pi v^2\int_0^\infty dr \, r^2\left[\phi^{\prime \,2}+\frac{1}{4}%
\,m_H^2\,(\phi^2-1)^2 \right] 
\label{Higgs_Hamiltonian}
\end{equation}
where $m_H^2 \equiv 2 \lambda v^2$ is the Higgs physical mass. We will take $v=246\,GeV$ and $m_H=100\,GeV$ and use units of $100\,GeV$ throughout the paper. 

Consider now the addition of fields with a global internal symmetry, leading to a conserved particle number $N$ associated to each field.  If these $N$ particles (fermions or bosons), couple strongly to the Higgs field, or $N\gg1$, the Higgs  could be strongly distorted in the vicinity of these particles.  To describe this situation, we adopt a mean-field approximation, in which all the particles are described by the \textit{same} wave functions in the background of a non-trivial Higgs field. Corrections to this mean-field description, such as, many-body, recoil and retardation of the Higgs field are expected to be suppressed by factors $v/m$, $m_H/m$ and $1/N$. 
In the semiclassical approximation, the total energy of the system will  be given by (see derivations in the Appendix A for $W $and Appendix B for fermions)
\begin{equation}
E_{cl}=E_{Higgs} + \sum_{a}n_{a}\varepsilon _{a},
\label{Energy}
\end{equation}
where $\{\varepsilon_a\}$ is the spectrum of the corresponding field in the Higgs background,  $n_{a}$ is the occupation number of each state and $N=\sum_{a}n_{a}$ is the total, conserved, particle number.

In the Higgs vacuum, i.e. $\phi(r)=1$, the state of lowest energy with $N$ particles corresponds to these particles sitting at the bottom of the continuum spectrum, with total energy $NM$. However, in the background of a non-trivial Higgs field there are two competing effects. On the one hand, the gradient and potential terms in (\ref{Energy}) increase the energy but, on the other hand, there might be some bound states levels with energy $0 < \varepsilon_{a} < M$ which can allocate the quanta, lowering the energy of the system of particles at the expense of creating such distortion. Therefore, we consider (\ref{Energy}) as a functional of $\phi(r)$ and search for non-trivial bag solutions. 

Let us start by a crude estimate of the the order of magnitude of $N$ for which such bags may exist. If we were to deplete  a certain large volume of the Higgs VEV (surface/kinetic terms neglected for now), it would require an energy
\be{
 \textit{Vol} \cdot  \frac{m_H^2}{8} v^2.\nonumber
}\ee 
For a bag of radius, say, $R\cdot 100\, GeV$=4, this energy is about 20 $TeV$. Thus, if the lowest  W-boson energy level has a binding energy of the order of $30\,GeV$ per $W$ or $Z$, an order of $O(1000)$ of them would be needed to compensate for the bag energy and
obtain some binding. The top quarks are  heavier and may get much larger binding, so one might naively think that less of them would suffice.  But the situation with fermions is  more delicate due to the the Pauli exclusion principle: we will discuss this issue in detail below.

\subsection{ Bosons in a bag}
\label{sec_bags}

The bags in which Higgs VEV is modified nearly to zero in some volume --to be called the ``no-Higgs" bags below --
makes particles which are heavy in vacuum  nearly massless.
Such bags are analogous in spirit to the various bags used in the 1970's to model
 the QCD confinement (e.g. the famous MIT bag \cite{Chodos:1974je}), except of course there is no need to take
the  mass  outside the bag to infinity.
If the particles considered are gauge bosons, $W,Z$, they all occupy the lowest bound state level. The energy of a bag with radius $R$  is

\begin{equation}
E={\frac{C_R N}{R}}+C_{V} R^{3},
\end{equation}
where the first term is the kinetic energy of (approximately) massless quanta confined in the bag. The precise constants depend on the details of the model, one example being $C_R\approx 2.04$ for a fermion confined in the MIT bag. It is clear that in such a case there is always a minimum and minimization with respect to $R$, at constant $N$, gives the size $R(N)$ of the bag as a function of $N$. One finds that the 
total energy per particle $decreases$ with $N$ as $E/N\sim N^{-1/4}$, for particles with mass $M_i$,  the binding ($E/N<M$) occurs above the  critical number 
\be
N_c=\left(\frac{4}{M_i}\right)^4\left(\frac{ C_V^{1/3}C_R}{3}\right)^3
\ee 
 Note also that even if the bags are not absolutely bound, there might exist metastable bags, local minima with $E/N>M$,  for $N_{\ast}<N<N_c$,
while  for $N<N_{\ast}$ 
   bosonic bags do not exist at all.

There is another, maybe less obvious, possibility of the bag structure, which we shall refer to as the ``inverted bag". In this configuration, we let the Higgs reverse its sign inside the bag, crossing zero at some finite radius $r_0$, and asymptotically taking the usual Higgs VEV $v$. If the Higgs field comes close to $-v$ inside the sphere, such a bag is simply filled with another Higgs vacuum, so
there is no volume energy but only a surface contribution.  It is reasonable to expect (and we shall see later that it is indeed the case) that  particles will be localized around the Higgs' node since they are massless there. If so, a large number of them  would form a 2-dimensional gas moving approximately on a sphere surrounding  an opposite-sign vacuum. 


\subsection{Fermions in a bag}

Let us start with a ``no-Higgs" bag picture.
As in any macroscopic situation, both the energy of the gas and of the Higgs field scale as the volume and the (mechanical) equilibrium in such case is achieved when the internal pressure due to fermions is compensated by the external pressure of the Higgs vacuum. It is easy to see, on dimensional grounds, that for massless fermions the pressure scales as $p_{f}\sim \mu^4$ and hence $N=\partial E/\partial \mu \sim \mu^3 R^3$. 
Therefore, $\mu \sim N^{1/3}/R$, and hence the total energy reads

\begin{equation}
E={\frac{A_V N^{4/3}}{R}}+C_{V} R^{3}.
\label{volumebag}
\end{equation}%
Minimizing the energy over $R$, with fixed $N$, provides the pressure balance condition mentioned above, leads to
\be{
 \frac{E}{N}=4\left(\frac{C_V^{1/3} A_V}{3}\right)^{3/4}.
}\ee
The binding $E/N<M_i$, contrary to bosonic bags, gives a condition which is \textit{independent of $N$} and whether binding will occur or not will depend exclusively on the values of the coupling constants and Higgs VEV, for any  number of particles.

Another option, an ``inverted bag", would put fermions  at the Higgs VEV node at $r\approx r_0$, within a shell of the width $O(1/m_H)$. Therefore, for large $N$ we expect  the system to be described by a 2-dimensional fermionic gas of volume $\propto r_0^2$.  For a 2D gas the pressure scales as $p_{f}\sim \mu^{3}$ and hence $N\sim \mu^2 r_0^2$ and therefore the total energy is given by
\begin{equation}
E={\frac{A_{S}N^{3/2}}{r_0}}+C_{S} r_0^{2}
\end{equation}%
with two other constants $A_{S},C_{S}$.
The energy per particle 
\begin{equation}
\frac{E}{N}=3\, \left(\frac{C_S^{1/2} A_S}{2}\right)^{2/3}
\label{EnergyParticleSurface} 
\end{equation}
is again saturated (N-independent). So whether binding occurs or not depends on the value of couplings and Higgs VEV. The possible existence of the configuration  may seem less intuitive, but  a closer look at our system reveals an interesting situation,  which supports this picture 
for fermions more than for bosons. For very large bags of this kind, the system becomes effectively 1-dimensional (see Appendix  \ref{Appendix_fermions}) and the Higgs equation (\ref{Higgs_equation_motion}) admits a 1-dimensional kink solution, located at $r=r_0>>1/m_H$. It is well known that
the Dirac spectrum in this background contains a series of discrete fermionic levels, including a \emph{fermionic zero mode}, so in this limit the localization of (at least some) fermions on the (large sphere) surface costs no energy at all!  

\section{Gauge boson Bags}
\label{sec_bosons}

    \subsection{General equations for E,  L and M modes}
    \label{static_field}
    
Consider the propagation of $W$-bosons in an external Higgs field as described by the  electroweak sector of the SM Lagrangian density (see Appendix \ref{Appendix_bosons})    
     \begin{eqnarray}
      \nonumber
      {\mathcal L} ={\mathcal L_{Higgs}} -\frac{1}{2}\left|\partial_\mu W_\nu -
        \partial_\nu W_\mu\right|^2      
        +M_W^2\phi^2 \,\left|W_\mu\right|^2
          \label{W1}   
    \end{eqnarray}
Due to the remaining global symmetry $W_{\nu} \rightarrow e^{i \alpha} W_{\nu}$, there's an associated conserved current given by
\begin{IEEEeqnarray}{rCL}
j^{\mu }&=&i\Big[ W_{\nu }\left( \partial ^{\mu }W^{\nu \ast }-\partial ^{\nu
}W^{\mu \ast }\right) \\ \nonumber 
&&-W_{\nu }^{\ast }\left( \partial ^{\mu }W^{\nu
}-\partial ^{\nu }W^{\mu }\right)\Big] 
\end{IEEEeqnarray}
with the associated conserved charge 
   \begin{equation}
N=\int d^{3} xj^{0}.
\end{equation}
 The classical equations of motion read
    \begin{equation}
      \label{wave}
      \left( \Box+M_W^2\phi^2\right) W^\mu 
     - \partial^\mu \partial^\nu \,W_\nu =0~.
    \end{equation}  
    
By taking a derivative of Eq. (\ref{W1}),  one can rewrite the equation of motion in the more transparent form
   \begin{eqnarray}
      \label{form}
      \left( \Box+M_W^2\phi^2\right) W^\mu
     +\partial^\mu \left(\frac{ W^\nu\partial_\nu \phi^2}{\phi^2}\right) =0.~~\quad
    \end{eqnarray}
 It will prove convenient to study these equations of motion in the electric (e), longitudinal (l)  and magnetic (m)  basis (see, e.g., \cite{LL4}). Assuming stationary fields and spherical symmetry, we write  
\begin{eqnarray}
\boldsymbol{W}^{(e,l,m)} & =& \boldsymbol{ Y}^{(e,l,m)}_{jm} f_{(e,l,m)}(r)/r, 
\end{eqnarray}
where $\boldsymbol{ Y}_{jm} $ are spherical harmonic vectors and $ f_e(r), f_l(r)$ and $f_m(r)$ are the radial wave functions for each mode. As seen in Appendix \ref{Appendix_bosons}, in a static, spherically symmetric, background the last term in (\ref{form}) vanishes for the magnetic mode, leading to the simple Klein-Gordon equation  for the radial wavefunction
       \begin{eqnarray}
      \label{magnetic}
\left(\frac{d^2}{dr^2} + \omega^2 - M_W^2\phi^2 -\frac{j(j+1)}{r^2} \right) f_m(r)
=0
    \end{eqnarray}
  for which $j\geq1$. However, the Laplacian mixes the electro-longitudinal modes, leading to the set of coupled equations
       \begin{eqnarray}
      \label{electric}
\left( \frac{d^2}{dr^2} + \omega^2 - M_W^2\phi^2 -\frac{j(j+1)}{r^2} \right) f_e(r)+ \\ \nonumber
-\frac{2\sqrt{j(j+1)}(1-r\frac{d \ln{\phi}}{dr})f_l(r)}{r^2}=0.
    \end{eqnarray}

      \begin{eqnarray}
      \label{longitudinal}
\left(  \frac{d^2}{dr^2} + \omega^2 - M_W^2\phi^2 -\frac{j(j+1)}{r^2} \right) f_l(r) \\ \nonumber 
+ 2r{d \over dr} \left( {f_L(r) \over r} {d \ln \phi \over dr} \right) -
\frac{2\sqrt{j(j+1)}f_e(r)}{r^2}=0. 
  \end{eqnarray}

While these equations have similarities with the Klein-Gordon and Dirac equations (if the latter one is written at the second-order differential equation), but there is also an important distinction.  The last term in (\ref{form}), which is not present for scalars or spinors,  becomes singular in the region where the Higgs field  $\phi$ crosses zero, effectively repelling the longitudinal modes from the node (or the whole bag, in the case of
the ``no-Higgs" bag). It is qualitatively explained by the fact that massless gauge fields have no longitudinal mode at all. 
As a consequence, below we will only discuss magnetic modes, which are unaffected by mixing with the longitudinal one.  
 
\subsection{Spectrum} 
 
 \subsubsection{Analytical Results}
 
 As mentioned before, the magnetic modes are not mixed with the longitudinal W boson modes which get
 repelled from the bags, so  we study the spectrum of (\ref{magnetic}) delegated studies of the 
 $j=0$ longitudinal mode in the Appendix. We shall separately study the cases of an inverted and no-Higgs bag. 
 
 In the case of a bag with no node, assuming that the Higgs takes some constant value $\phi_0$ with $0<\phi_0<1$ for $r<R$, and $\phi=1$  for $r>R$, the solutions to  (\ref{magnetic}) are given by spherical Bessel functions with frequency squared $\omega^2-M_W^2 \phi_0^2$. Assuming that the wavefunction is regular at the origin and vanishes at $r=R$, the spectrum for the lowest radial excitation of the magnetic mode is given by 
 
 \begin{equation}
 \omega^2=\left( \frac{ \beta_{1, j+1/2}}{R}\right)^2+M_W^2 \phi_0^2,
 \label{BesselOmega}
 \end{equation} 
 where $\beta_{1, j+1/2}$ is the 1'st zero of the Bessel function and $j\geq 1$.  Therefore, bound magnetic W-bosons will fill the whole of the bag, occupying the lowest level with an energy given by (\ref{BesselOmega}).

In the case of an inverted bag, we assume that  $\phi(r)$ has a node at large $r_0$, i.e. $\phi(r_0)=0$, with $\phi^{\prime}(r_0)>0$. We expand around $x=r-r_0=0$ and Eq. (\ref{magnetic})  is a simple harmonic oscillator
 
 \begin{equation}
 f_m^{\prime \prime} +\left[ \left( \omega^2-\frac{j(j+1)}{r_0^2} \right) - M_W^2\phi^{\prime \, 2}(r_0)x^2\right] f_m =0
 \end{equation}
 and hence we find
 \begin{equation}
 \varepsilon_0^2 \equiv \omega^2=  M_W\phi^{\prime}(r_0)+\frac{j(j+1)}{r_0^2}  \label{WLevels}
 \end{equation}
 for the lowest level. Therefore, for large bags with a node, the wave functions of magnetic W-modes  are Gaussian and localized around the Higgs's node, with energy eigenvalues given by (\ref{WLevels}).

 \subsubsection{Numerical Results}
 
\begin{figure}[t!]
\includegraphics[width=2.7in]{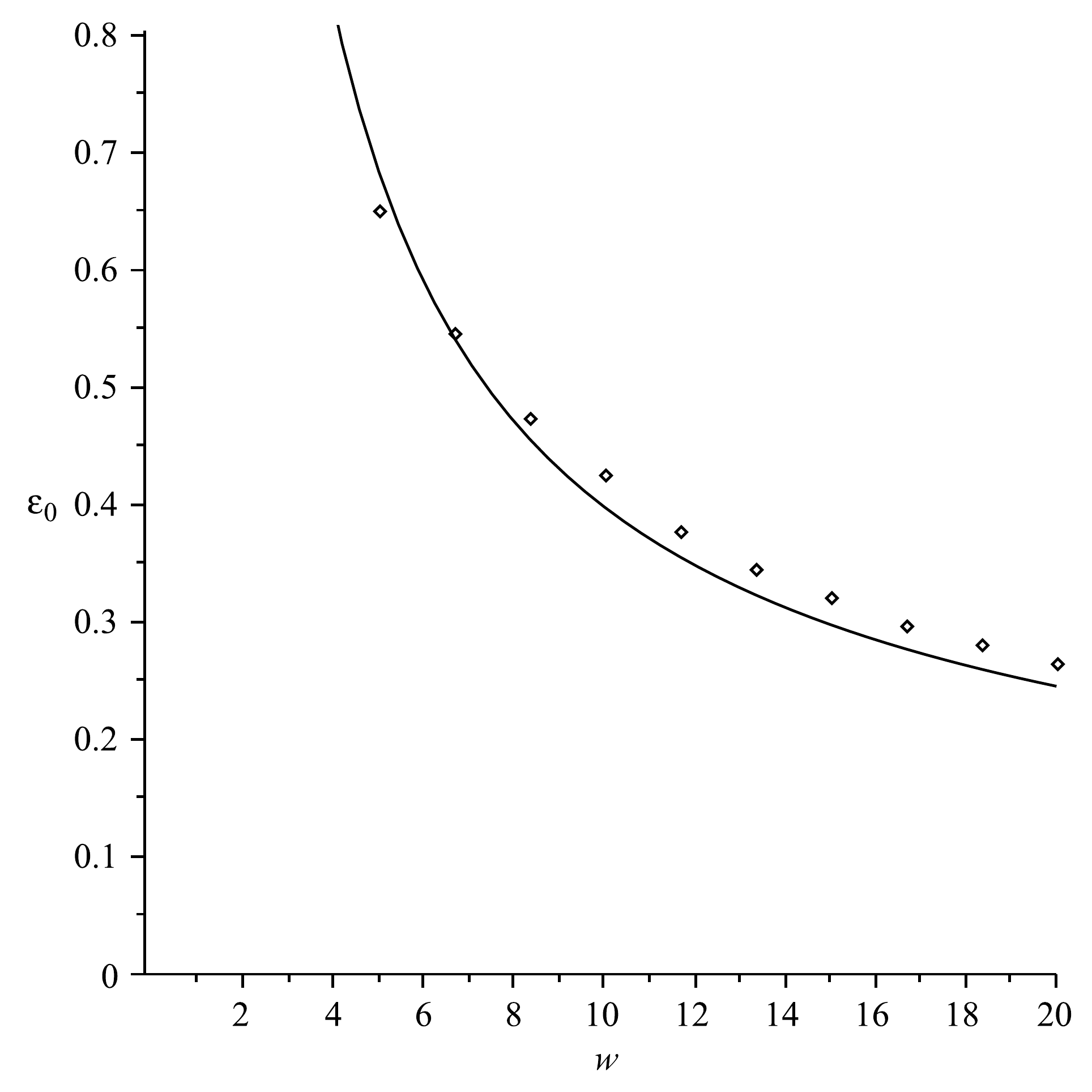}
\caption{Lowest magnetic mode ($j=1$) levels for a Gaussian bag with $\alpha=1.3$. The numerical results (points) are in good accordance with the analytical result (\ref{WInvertedLevel}). }
\label{W_M_Levels_a13}
\end{figure}

In order to find bag solutions for finite $N$, we adopted a variational approach and took as  a trial function for the Higgs, the Gaussian profile

 \be
  \phi(r)=1-\alpha \, \exp{(-r^2/w^2)},  \label{eqn_Gaussian} 
  \ee 
where the two parameters, $\alpha$ and $w$ describe the depth and the width of the bag potential, respectively. Solving the W-boson magnetic equation (\ref{magnetic}) in this Higgs background is rather straightforward using a shooting method, and we may compare the results to the analytical expressions for large-size bags derived previously. For instance,  for the case of an inverted bag (i.e.  $\alpha>1$), the Higgs has a node at $r_0=w \sqrt{\ln \alpha} $ and from (\ref{WLevels}) we find

\begin{equation}
 \varepsilon_0=\left( \frac{2 M_W \sqrt{ \ln{\alpha}}}{w}+\frac{2}{w^2  \ln \alpha } \right)^{1/2}
 \label{WInvertedLevel}
\end{equation}
for the lowest magnetic mode.
In Fig. \ref{W_M_Levels_a13} we compare the numerical with this analytical result for a bag with $\alpha=1.3$ and find good agreement. Similarly, we also found  numerically the spectrum for a ``no-Higgs" bag with $\alpha=1$.
 
\subsection{Existence of pure W-bags}

As shown in Appendix \ref{Appendix_bosons}, the total energy of the system with $N$ W-bosons in the lowest energy level $\varepsilon_0$ is
\begin{equation}
E_{cl}=E_{Higgs}+N\varepsilon_0.
\end{equation}
Using a variational approach based on the Higgs ansatz (\ref{eqn_Gaussian}),  we calculate the Higgs energy (\ref{Higgs_Hamiltonian})  in terms of $\alpha$ and $w$
\begin{multline}  E_{Higgs}=\frac{ (2 \pi \alpha v)^2}{16 \sqrt{2 \pi}}\left. \Big[ 3 w  \right.  \\ 
+ \left. w^3 M_H^2\left(1 -\frac{4}{ 3\sqrt{6}} \alpha +\frac{1}{8 \sqrt{2}} \alpha^2\right)\right]. 
\label{eqn_Higgs_energy}
\end{multline}
and search for the local minima in this 2-dimensional parameter space.  Since the behavior of bags with $\alpha>1$ and $\alpha<1$ is different, we present these cases separately. 

\subsubsection{The no-Higgs Bag}

\begin{figure}[t!]
\includegraphics[width=2.7in]{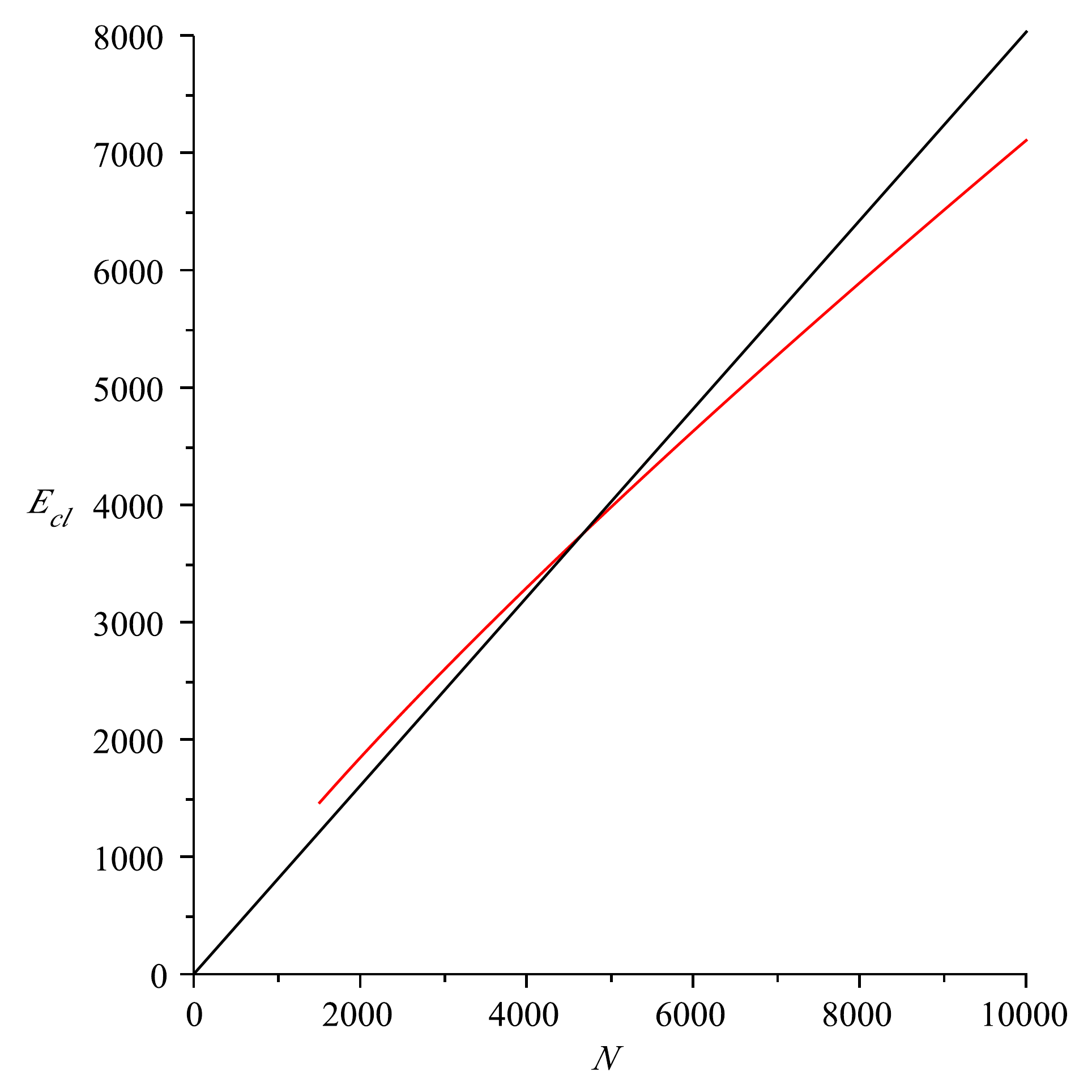}
\caption{ Total energy/$100\,GeV$ of bag  with $\alpha=1$ (red curved line) and that of simple plane waves (black straight line) as a function of $N$. For $N>N_c \sim 4700$  the bags are stable, while for $N_c>N>N_{\ast}\sim 1400$, bags exist only as metastable states. For $N<N_{\ast}$ bags do not exist at all. 
}
\label{W_bag_N}
\end{figure}

For our cosmological applications it would be of particular importance to have near-empty bags with  $\alpha \approx 1$,
because only there the electroweak sphalerons can be represented by pure-gauge (COS) sphalerons  which have larger probability.
So, in order to demonstrate this situation, we took $\alpha=1$, solved the W-boson spectrum numerically and proceeded to calculate the total energy of a bag with $N$ quanta. In Fig. \ref{W_bag_N} we show the total energy of bag  with $\alpha=1$ (red curved line) and that of simple plane waves (black straight line), as a function of $N$. For $N>N_c \sim 4700$  the bags are energetically favored and thus stable, while for $N_c>N>N_{\ast}\sim 1400$, bags exist only as metastable states. For $N<N_{\ast}$ the bag would be to small to hold any levels and cannot exist at all. As one can see from this figure, the general feature of the bags in question is that the overall energy per particle
is not very different from the particle mass: and yet the way this is reached is completely different from individual particles without a bag, since the quanta are 
near-massless inside the bag. 

\subsubsection{Inverted Bag}

\begin{figure}[t!]
\includegraphics[width=2.9in]{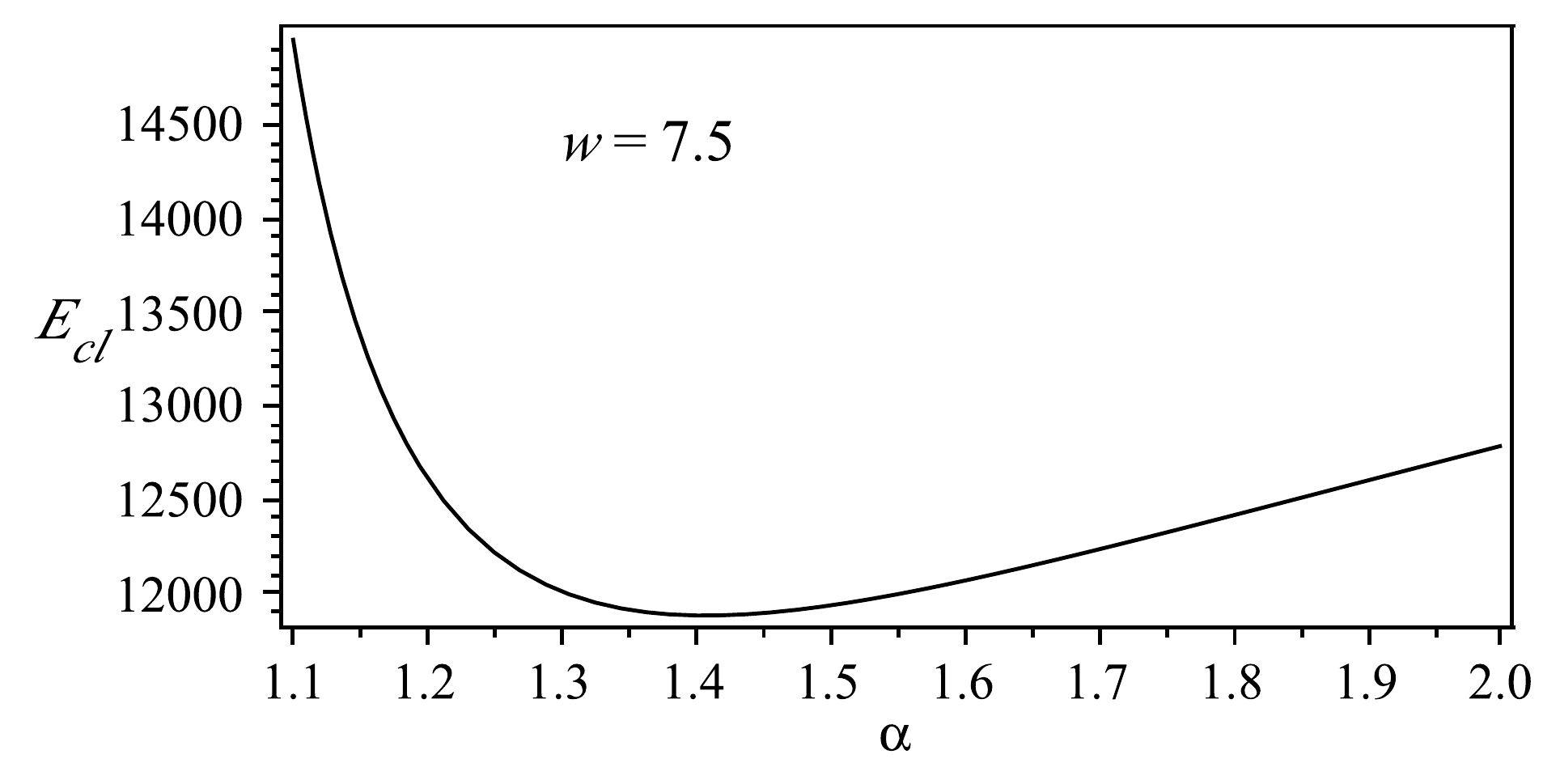}\\
\includegraphics[width=2.9in]{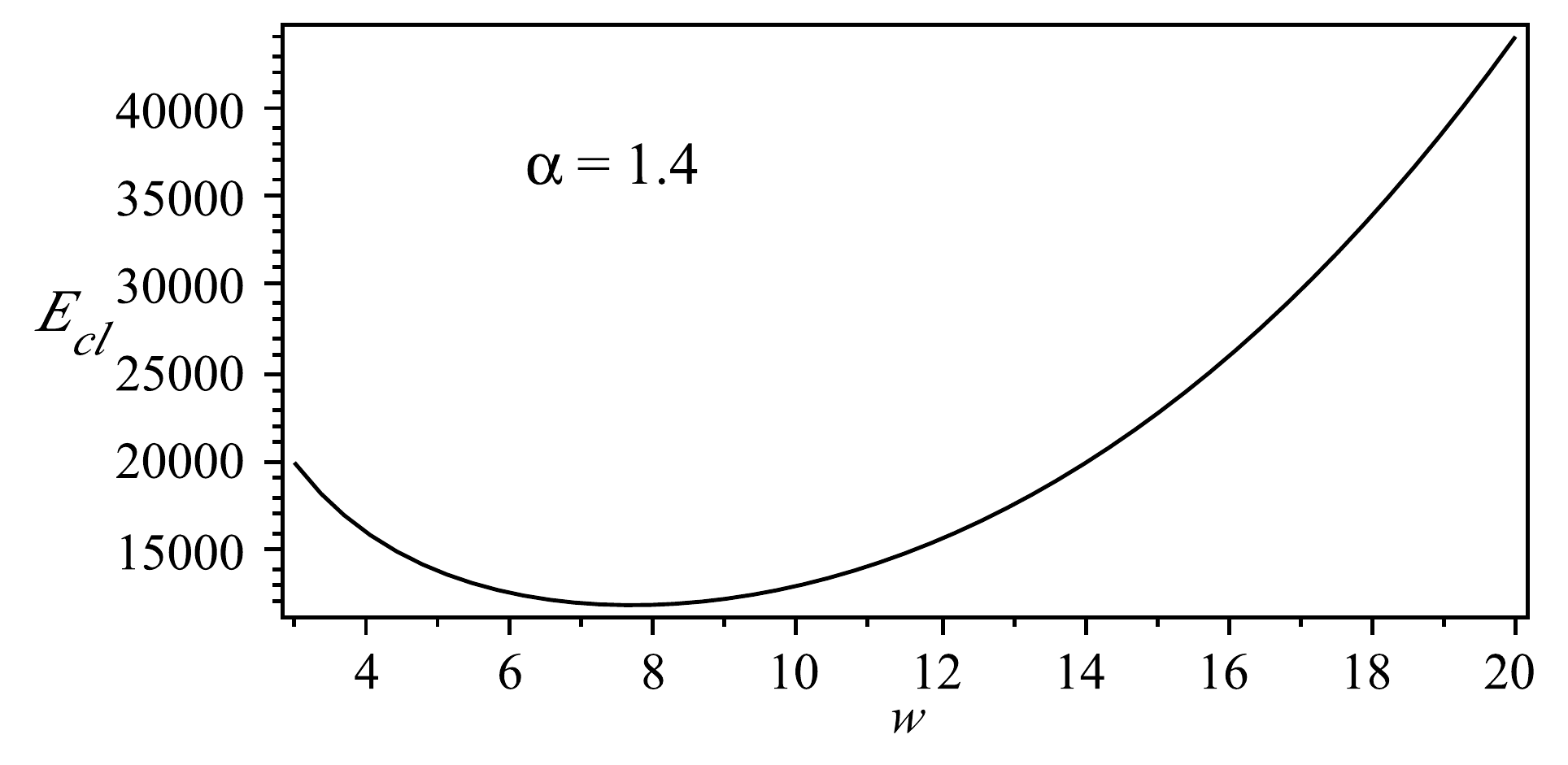}
\caption{Total energy/$100\,GeV$ of an inverted-type W bag with $N=20,000$ in the lowest magnetic mode in terms as a function of $\alpha$ and $w\cdot100\, GeV$. In this example, a local minimum  in both directions  is found for $\alpha=1.4$ and $w=7.5$.}
\label{3DParameterWBag}
\end{figure}

From Eqs. (\ref{WInvertedLevel}, \ref{eqn_Higgs_energy}) we also searched for inverted-type ($\alpha>1$)  bag solutions by looking for local minima of the total energy in both the $\alpha$ and $w$ directions at fixed $N$ . The binding of such bags occurs for a rather larger $N$, but do occur. In Fig. \ref{3DParameterWBag} we show an example for  $N=20,000$ for which a local minimum was found at  $\alpha=1.4$ and $w=7.5$.

\section{Fermion bags}
\label{sec_fermions} 
\subsection{The Dirac equation}

Omitting for now gauge bosons, we consider  a system of $N$ heavy fermions interacting with a background Higgs field $\Phi$, as described by the SM Lagrangian density

\begin{equation}
\mathcal{L} =\mathcal{L}_{Higgs}+\bar{\psi}\left( i\gamma ^{\mu }\partial _{\mu }-g\Phi\right) \psi,
\label{lagrangian}
\end{equation}%
where $\Phi $ is complex scalar in the fundamental representation of $SU(2)_{L}$ and $\psi$ is a Dirac spinor in 3+1 dimensions. Due to the global symmetry $\psi \rightarrow e^{i \alpha} \psi$, there's an associated conserved charge given by

\begin{equation}
N \equiv \int d^3x \psi^{\dagger} i \partial _{0 }\psi ,
\end{equation}
which is identified with the fermion number. For completeness, let us remind the reader the standard notations for Dirac
spinors in spherical coordinates \cite{Landau}:

\begin{equation}
\psi =\frac{1}{r}\left( 
\begin{array}{c}
F(r)\Omega _{jlm} \\ 
(-1)^{1/2(1+l-l^{\prime })}G(r)\Omega _{jl^{\prime }m}%
\end{array}%
\right),
\end{equation}%
where $\Omega _{jlm}$ are spherical 2-component spinors and we take normalization $\int dr \,(F^{2}+G^{2})=1$. The so-called Dirac parameter $\kappa$ is defined as 
\begin{equation}
\kappa =\left\{ 
\begin{array}{c}
-(l+1)\text{ \ \ \ for }j=l+1/2 \\ 
l\text{ \ \ \ \ \ \ \ \ \ \ \ \ \ \ for\ }j=l-1/2%
\end{array}%
\right. 
\end{equation}
and runs over all nonzero integers, being positive for anti-parallel spin and negative for parallel spin. Dirac's equation reads (See Appendix \ref{Appendix_fermions}): 
\begin{eqnarray}
 (\varepsilon-m\,\phi)\,F~\, & = &\,-G^{\prime}\,+\,(\kappa/r)\,G \\ \nonumber
(\varepsilon+m\,\phi)\,G~\,& = & \,\,~~F^{\prime}\,+\,(\kappa/r)\ F  
\label{Dirac_Equation}
\end{eqnarray}
The form of these equations presumes that the eigenvalue $\varepsilon$ is positive. A negative eigenvalue would correspond to a state in the lower fermion continuum.  If so, a charge conjugation transformation turns it into a positive eigenvalue for an antifermion. The Higgs equation of motion reads
 \begin{equation}
\phi^{\prime\prime}+\frac{2}{r}\phi^{\prime}+\frac{m_H^2}{2}\phi\,(1-\phi^2)\,=\,\frac{(N-1)m}{4\pi \,v^2}\,\frac{F^2-G^2}{r^2}
\label{Higgs_equation_motion} 
\end{equation}
Note that in the $r \rightarrow \infty$ limit, the source term in the right hand side, as well as the additional $\phi^{\prime}/r$ term from the Laplacian can be neglected and the equation becomes the usual equation for a 1D kink.

\subsection{Spectrum}

\subsubsection{Analytical Results}
\label{Spectrum_Fermion_Analytical}

 In order to analytically justify the physical picture of inverted bags , we calculate the
energy of surface-fermions for an inverted bag of large (but finite) radius.  We assume that $%
\phi(r)$ has a node at large $r_0$,  at which the derivative is \textit{positive}, $\phi'(r_0)>0$,  we search for solutions of the Dirac equation in the near-Gaussian form 
\begin{eqnarray}
&F(r)\,=\,A(r) \,e^{-S(r)\,}~,  \label{FA} \\
&G(r)\,=\,B(r) \, e^{-S(r)\,}~, \nonumber  \label{GA} \\
S(r)\,&=\,m\,\int_{r_0}^r\phi(r^{\prime})dr^{\prime}\,\simeq\,\frac{1}{2}\,m
\,\phi^{\prime}(r_0)\,(r-r_0)^2~.  \nonumber  \label{S}
\end{eqnarray}
The large mass $m$ here favors the localization of $F(r)$ and $G(r)$ in the
vicinity of $r_0$, which justifies the approximation. The functions $A(r),B(r)$ can be iteratively found from
\begin{eqnarray}
\label{ABBA}
\begin{array}{l}
\varepsilon \,A+B^{\prime}-(\kappa/r)\,B\,=\,~~m \phi \,(A+B) ~  
\\
\varepsilon \,B-A^{\prime}-(\kappa/r)\,A\,=\,-m \phi \,(A+B) ~.  
\end{array}
\end{eqnarray}
Since $A, B$ are smooth functions of $r$ we expand to first order about $r=r_0$  and derive a system of four linear homogeneous algebraic equations for $(A_0, B_0, A_0^{\prime}, B_0^{\prime})$, given by the matrix

\begin{equation*}
\begin{pmatrix}
\varepsilon & - \frac{\kappa}{r_0} & 0 &1\\
- \frac{\kappa}{r_0} & \varepsilon & -1 & 0\\
-m \phi^{\prime}(r_0) & -m \phi^{\prime}(r_0) +\frac{\kappa}{r_0^2} & \varepsilon & -\frac{\kappa}{r_0} \\
 m \phi^{\prime}(r_0) +\frac{\kappa}{r_0^2}  & m \phi^{\prime}(r_0) &  -\frac{\kappa}{r_0} & \varepsilon 

\end{pmatrix}.
\end{equation*}
Setting $\det=0$, we solve for  $\varepsilon$ and get
\begin{equation}
\varepsilon_{\pm}^2=m \phi^{\prime}(r_0)+\kappa^2/r_0^2\pm \sqrt{m^2
\phi^{\prime 2}(r_0)+\frac{1}{r_0^4}}.
\label{epsilon_fermions}
\end{equation}
We see from this expression that the levels $\varepsilon_{+}$ are finite for $r\rightarrow \infty$, while $\varepsilon_{-}$ goes to zero. The latter correspond to the would-be zero modes in the 1 dimensional case. Therefore, for large enough bags these will be below any other level  (see Appendix \ref{Appendix_fermions} for explicit results) and in what follows we consider these levels only.  Note that to order $O(\frac{1}{ r_0})$,  
\begin{equation}
\varepsilon_{-}\,\simeq{\frac{|\kappa| }{r_0}},
\label{eps}
\end{equation}
hence in this approximation the spectrum is independent of the mass of the fermion and the value for the first level ($\kappa$=-1) is less than $2.04$, which appears for the volume bag. Thus, 
not only the Higgs energy, but the fermionic energy benefits as well. Note also that to this order, the functions $A, B$ are simply constants whose values are fixed by the normalization condition to
\begin{equation}
A\,=\,-\,B\,=\,[\,m\phi^{\prime}(r_0)/4\pi\,]^{1/4}~.  \label{ABC}
\end{equation}
Therefore, in this limit $F^2 \sim G^2$ and thus these
states do not contribute to the scalar source $F^2-G^2$ in the Higgs equation.
This is because this approximation corresponds to the 1-dimensional zero-mode solution ($F$ and $G$ being the upper and lower components of a 1+1 Dirac spinor), which
has a particular chirality (i.e. $F = \pm G$) and thus no scalar source. However, this is only
true for the leading terms: As $r_0$ becomes smaller, the $\sim 1/r_0$ terms of our expansion
violate this equality, rendering the scalar $F^2-G^2$ nonzero.


\subsubsection{Numerical results}

\begin{figure}[t!]
\includegraphics[width=2.7in]{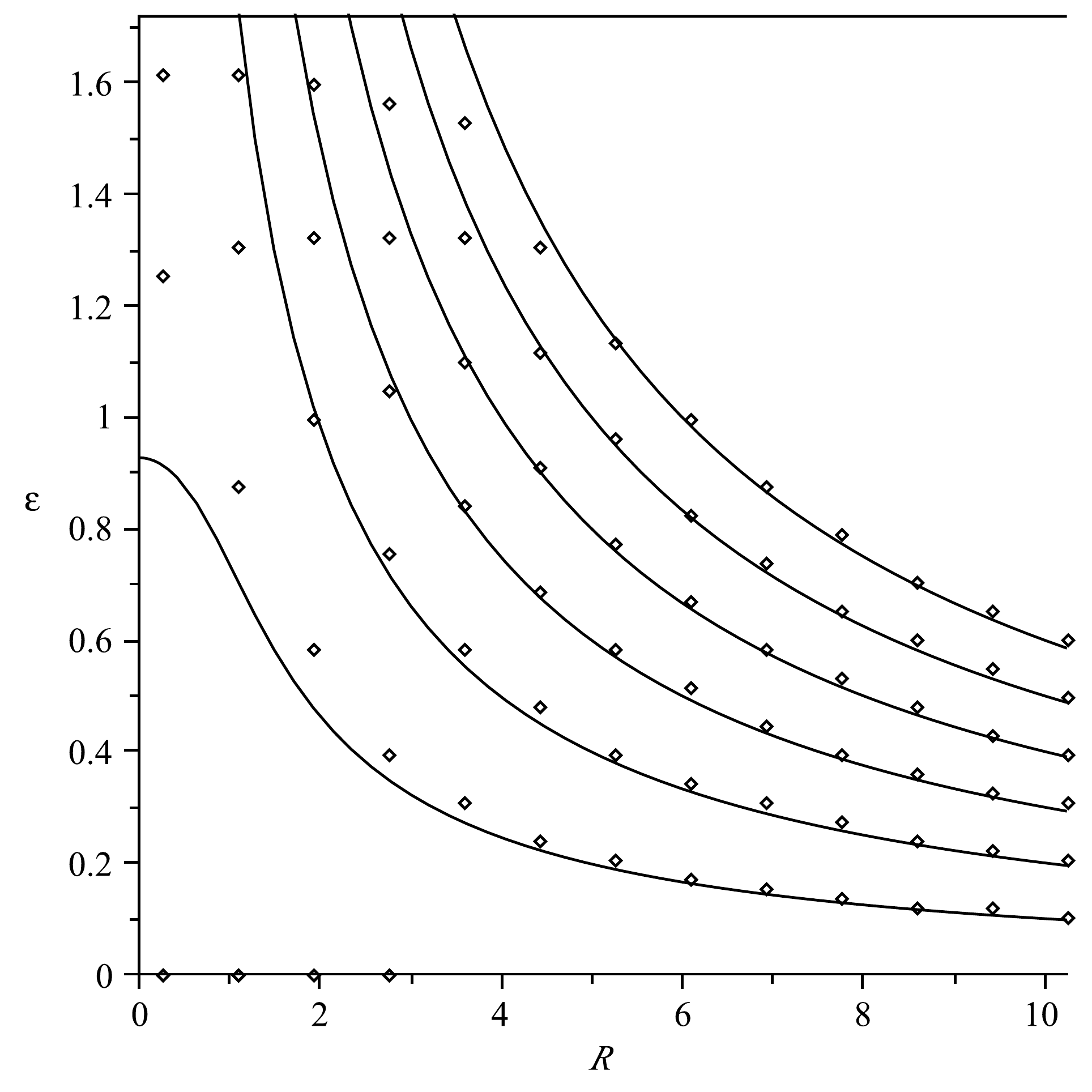}
\caption{Energy levels (in units of $100\, GeV$) of the Dirac operator in the background of an inverted bag ($\eta=1$) as a function of  the bags's size $R\cdot 100\,GeV$. These levels correspond to  increasing $l$ (from the bottom up) for $\kappa<0$ only. The analytical $\varepsilon_{-}$ results for large $R$ (solid lines) are in good accordance with the numerical results (points) for even rather small $R$.}
\label{bag_spectrum}
\end{figure}

In this section we provide some of  the numerical results, and comparison to the analytical results derived before,  for the the Dirac spectrum in a Higgs background. We took an ansatz for the Higgs which interpolates between the type of bags  discussed above, namely surface and volume bags, parametrized as
\begin{equation}
\phi(r)= (1-\eta) +\eta\tanh\left[\frac{r- R}{\Delta r}\right],
\label{HiggsAnsatz}
\end{equation}
where $0\leq\eta\leq1$ and $0\leq R < \infty$, with $\eta=0$ corresponding to the trivial Higgs vacuum, $\eta=1/2$ to a no-Higgs bag and $\eta=1$ to an inverted bag. 
 By using a shooting method, the spectrum for Dirac's equation is found numerically  and examples of the levels for an inverted bag with $\eta=1$ are shown in Fig. \ref{bag_spectrum}. This bag has a node at $r_0=R$. One can see that there is very good agreement between the analytical expression (\ref{epsilon_fermions}) and the numerical results (points). Although it has been derived for large $R$, the agreement is not there only for small $R<2$ which we will not discuss in this work. 

\begin{figure}[t!]
\includegraphics[width=2.8in]{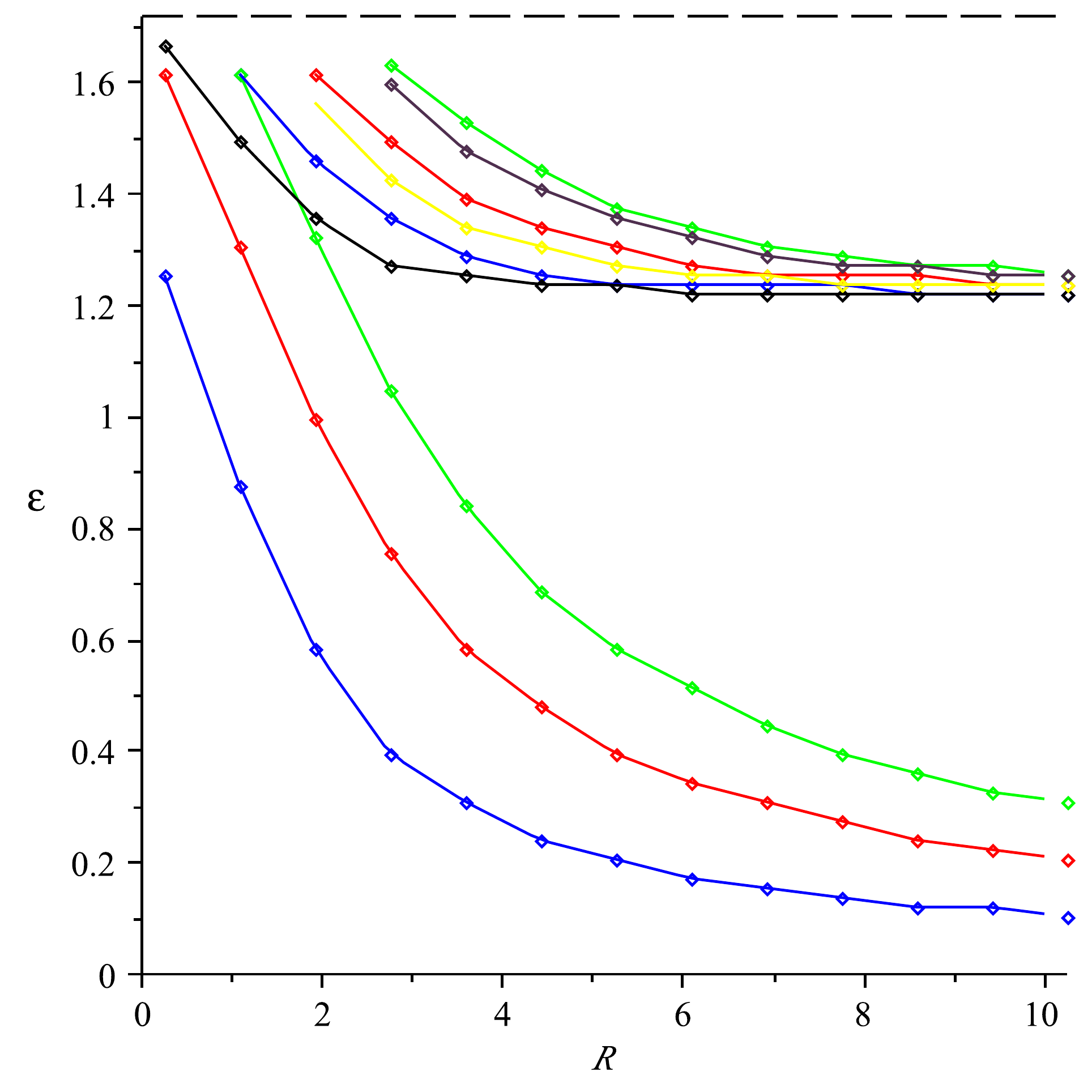}
\caption{(color online) Dependence of some  bound state levels energy $\varepsilon$ in units of $100\,GeV$ on the size of the bag, expressed as the parameter $R$ in units $1/100\, GeV^{-1}$ for a bag with $\eta=1$. The color coding of the levels and their quantum numbers are listed in Table I.}
\label{Top_Levels_Kink}
\end{figure}

\begin{table}[tbp]
\begin{tabular}{r|llllcl|c}
& $n_r$ & $\kappa$ & $l$ & $j$ & $Deg. (\bar t t)$ & color &  \\ \hline
1 & 0 & -1 & 0 & 1/2 & 12 & blue &  \\ 
2 & 0 & -2 & 1 & 3/2 & 24 & red &  \\ 
3 & 0 & -3 & 2 & 5/2 & 36 & green &  \\ 
4 & 0 & 1 & 1 & 1/2 & 12 & black &   \\ 
5 & 1 & -1 & 0 & 1/2 & 12 & blue &   \\ 
7 & 0 & 2 & 2 & 3/2 & 24 & yellow &  \\ 
8 & 1 & -2 & 2 & 3/2 & 24 & red & \\
9 & 0 & 3 & 3 & 5/2 & 36 & violet & \\
10 & 1 & -3 & 2 & 5/2 & 36 & green & 
\end{tabular}%
\caption{The properties of some levels, including the number of
radial nodes $n_r$, Dirac parameter $\protect\kappa$, orbital momentum $l$,
total angular momentum $j$, multiplicity of states $Deg$(for $t \bar t$ bags) and color code used in
our figures. Obviously, for pure top bags the multiplicity is halved.}
\end{table}

The Table shows magic numbers and the order in which levels are populated. Some levels  are also shown in
  Fig. \ref{Top_Levels_Kink} we show some levels (up to $l=2$).  As the bag's size decreases (or for smaller $\eta$), some levels with $\kappa>0$, and radial excitations of $\kappa<0$ states, cross and one must carefully fill the levels accordingly. Note that in this case, the lowest levels with $\kappa<0$ asymptotically become zero modes (with the correct chirality), while radial excitations thereof and levels with $\kappa>0$ merge, asymptotically becoming the next fermionic discrete level.

\subsection{Non-existence of pure top bags in vacuum}

\subsubsection{No-Higgs Bag}

Consider as a trial configuration a spherical cavity of radius $R$ with constant zero Higgs VEV inside, a small transition region of size $1/m_H$, and a constant Higgs VEV outside, just as in the Friedberg et al. bags, except that we shall occupy not only the lowest level. Note that this is a solution to (\ref{Higgs_equation_motion}), except in the (small) transition region. This will correspond to a free gas of massless fermions in a spherical volume of radius $R$ and statistical mechanics tells us that
\begin{equation}
E_{\psi}=\left(6 \pi/D\right)^{1/3}\left(3/4\right)^{4/3}\frac{N^{4/3}}{R},
\end{equation}
with $D$ the degeneracy factor due to spin, color and particle/antiparticle. From here we read the coefficient $A_V$ in (\ref{volumebag}). Directly from (\ref{Energy}) we also find $C_V=\frac{1}{6}\pi\,v^2\,m_H^2$ and therefore the energy per particle reads
\begin{equation}
\frac{E}{N}=\left(\frac{3 \pi^2 v^2 m_H^2}{D} \right)^{1/4}.
\label{Energy_per_particle_Empty}
\end{equation}
The bag will be bound provided $E/N< m_t$. Clearly, having a larger number of fermionic species helps: say $t+\bar t$ bags, with $D=12$, are
more economic than pure top bags with $D=6$.  From this expression (assuming $v=246\,GeV$) we find that the bag will be bound provided that $m_H<77 \, GeV$ which, unfortunately, is not the case in vacuum.

\subsubsection{Inverted Bag}

Given the  semiclassical solution derived in \ref{Spectrum_Fermion_Analytical}, we can estimate the binding energy for very large inverted bags. Finite-size bags require a detailed study of all the fermionic levels, including radial excitations and $\kappa>0$ states. However, as seen in Appendix  \ref{Appendix_fermions}, for finite but large enough bags the lowest levels are those given by (\ref{eps}) and therefore we shall fill these levels only for now. The maximum occupancy of each level is given by $n_{\kappa}=D|\kappa|$, with $D=6,12$ for only $t$ or $t\bar t$ bags, respectively. Therefore, for a large size bag the number of levels that it can hold, $\kappa_{max} \sim m r_0$,  is large and from (\ref{eps}) the fermionic energy reads
\begin{equation}
E_{\psi}\simeq \frac{D \sum_{\kappa=1}^{\kappa_{max}} \kappa^2}{r_0} \simeq \frac{D}{3} \frac{(2N)^{3/2}}{r_0}.
\end{equation}
where we have used  $\sum \kappa^2 \simeq \kappa^3/3$ and  $N=D \sum |\kappa| \simeq D \kappa^2/2$.
Thus, in the notations used for our generic estimates of the ``surface bag"
before, we have explicitly justified the 2-d Fermi-gas shell picture and, furthermore, found $A_S=D^{-1/2}\,2^{3/2}/3$. For large $r$,  equation (\ref{Higgs_equation_motion}) is solved by the kink profile
\begin{equation}
\phi(r)= \tanh[(r-r_0)m_H/2],
\label{kinksolution}
\end{equation}
which leads to $C_S=8.38\,m_H v^2$. These values for $A_S,C_S$ in (\ref{EnergyParticleSurface})
lead to an energy per-particle of 

\begin{equation}
\frac{E}{N}\simeq 3.68 \left(\frac{m_H v^2}{D}\right)^{1/3}.
\label{BindingEnergyLargeBags}
\end{equation}
Again, we see that binding in principle is possible provided the values of the parameters are such that $E/N< m_t$. However, this requires the even more restricting upper bound $m_H<20 \, GeV$, which is excluded experimentally and hence no binding of inverted top bags occurs.
\\

These results indicate that, for realistic Higgs mass,  pure top bags are not energetically favored over free tops. However, one may relax this condition and still wonder if at least \textit{metastable} top bags exist. To answer this we adopted a variational approach using the trial profile (\ref{HiggsAnsatz}) and searched for local minima in $(\eta, \, R)$. In the limit $R\gg \Delta r$ and using (\ref{Higgs_Hamiltonian}),  we  write the total energy in terms of $\eta, \, R$ and $\Delta R$. From here we find that the optimal value for  $\Delta r$ is $\Delta r^{(opt)}=\frac{2}{m_H (2-\eta)}$, which leads to the expression

\begin{multline}
\frac{E_{Higgs}}{2 \pi v^2 }=R^2 m_H \eta^2 \left( 2-\eta \right)\\
+ \frac{4}{3} R^3 m_H^2 \eta^2 \left(1-\eta\right)^2+...
\end{multline}
For $\eta=\{0,1\}$, corresponding to the usual Higgs vacuum and inverted bag, the volume term vanishes. If $\eta>1/2$ the bag has a node at $r_0=R+(2/m_H) \tan^{-1}[ (\eta-1)/\eta]$ and from from (\ref{epsilon_fermions}) we can parametrize the fermionic energy in terms of $(\eta, \,R)$ as well, which allows us to search for local minima in the 2-dimensional parameter space.
Although an interesting spectroscopy of these objects are found for small Higgs masses (e.g. $m_H\sim 50 \, GeV$), for a realistic Higgs mass, no local minima were found for any $N$. Therefore, no pure top bag solutions, stable or metastable,  were found in our calculations.

\section{Top quarks in W-bags}
\label{Mixed}
Although we have seen that pure top quarks are not heavy enough to create top bags by themselves (in vacuum), W-bags containing $N>N_c$ W-bosons certainly do exist and given that top quarks are present, one should address how is it that fermions behave in the presence of these bags. From the results of the previous section, we know that fermions  will bind to these bags, forming a 2-d fermi gas in the surface of an inverted bag or, filling the volume of a no-Higgs bag. The next issue to study is the \textit{relative} position of the W and top levels. 

Free (or weakly bound) top quarks are much heavier than $W$ bosons and thus decay
into another quark and $W$. In the presence of a large Higgs bag, however, the relative position of $W$ and top levels can be such
 that quark levels are lower than gauge boson ones. If this is the case, the life-time of $t$-quarks in such bags will be much longer, given by
 next order decays into three fermions, like in the usual beta decays.
 
\begin{figure}[t!]
\includegraphics[width=2.8in]{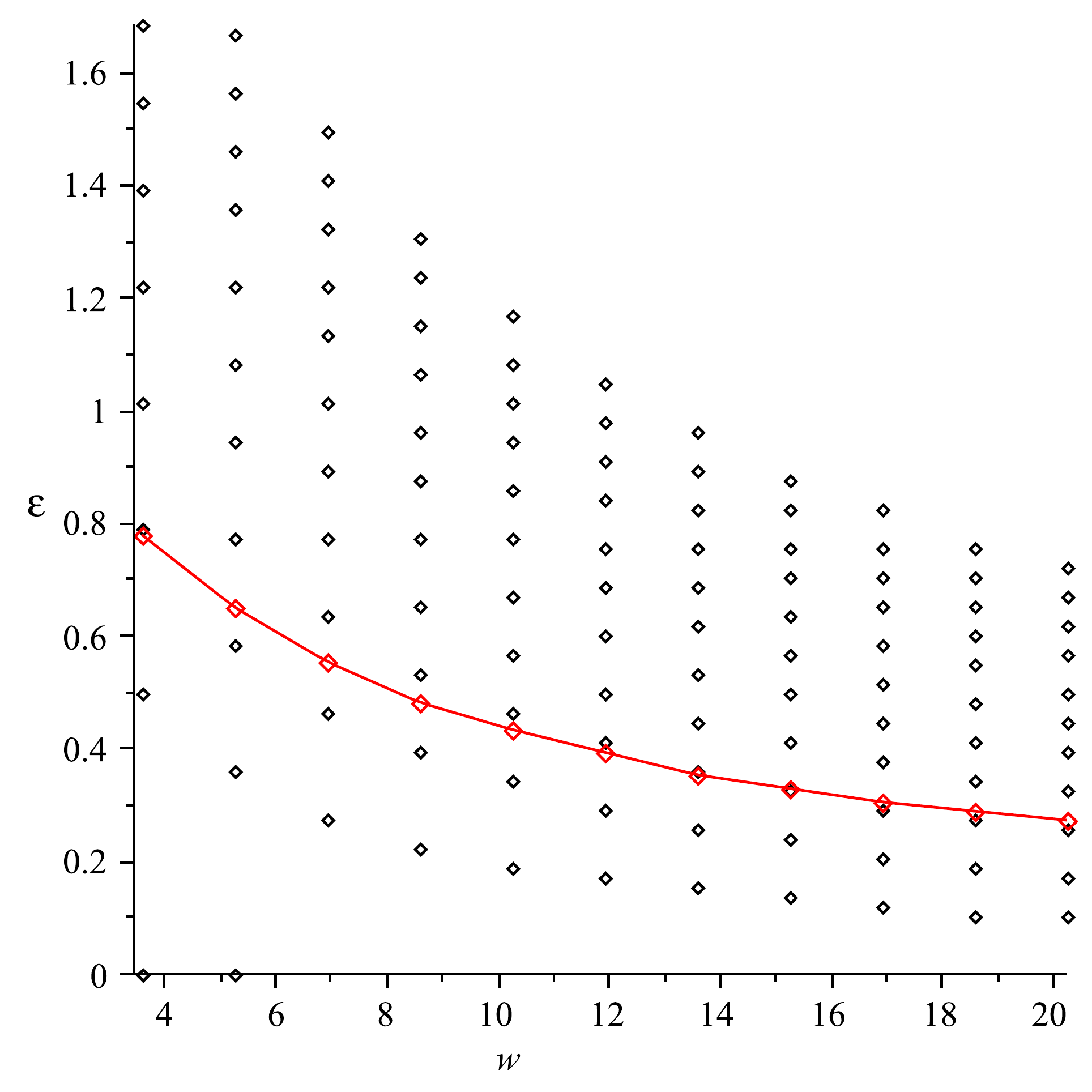}
\label{Mixed_bag_a=12}
\caption{The energy of the  levels $E_W/100 \,GeV$ versus the size of the bag
$w\cdot 100\, GeV$, for the Gaussian
ansatz for the Higgs and $\alpha=1.2$ (with zero case). Black (small) circles are some fermionic levels with $\kappa<0$ and the large (red) circles correspond to the W magnetic level for $ j=1$.
}
\label{fig_levels12}
\end{figure}

  To answer when this happens we numerically studied the behavior of these levels as a function of the bag's size. In Fig. \ref{fig_levels12} we show the resulting energy levels for the M-mode (red points/line) and  the top quark levels (small black circles) in the background of a Higgs bag with $\alpha=1.2$. 
Indeed, it is observed that when the bag becomes large the lowest-lying quark levels are $below$ the lowest $W$ ones, preventing their decay in the lowest order of weak interactions (to $Wb$).  We  discuss the possible cosmological role of these observations elsewhere \cite{FS_baryo}.

\section{Finite Temperature}
\label{sec_T}

As it has been already emphasized in the Introduction, these multi-quanta bags can hardly be produced experimentally, and their main application is
 cosmological, as  ``doorway states" facilitating  electroweak sphaleron rates as well as possible CP violation, hopefully to result in a possible mechanism for baryogenesis. In general, the possible scenarios can be divided into (i) Standard Cosmology, with very small deviations from thermal equilibrium and (ii) hybrid models, in which the electroweak phenomena are coincident  with the end of inflation.
 
 Although our paper \cite{FS_baryo}  is devoted to the latter scenario, with strong deviations from equilibrium, in this section we will restrict ourselves to the thermal equilibrium case, as a necessary intermediate step, extensively studied in the literature. For an overview and references see, e.g., \cite{Kapusta}.

  Before we come to specifics, let us outline the main issues to be discussed in this section. As the temperature  in the early universe 
  is ``at the electroweak scale", with momenta of the order of 100 $GeV$ or so, all parameters of the SM are significantly renormalized. This in particular refers to the Higgs mass $M_H(T)$ 
  and its VEV $v(T)$, which are strongly reduced as $T$ approaches the critical temperature $T_c$. 
  
  Recall that, as emphasized many times above, if the Higgs mass would be smaller, it would significantly increase binding of the multi-quark bags. Therefore, one may naively think that a high temperature environment will foster the existence of W-top bags. However, the part of the
  top and W masses related to Higgs are also reduced,  and so is their binding to the bag. On the other hand, only a part of the particles's masses come from Higgs mechanism: at finite $T$ they are complemented by the so called ``thermal masses" due to particle rescatterings. The outcome of these two effects is the subject of this section.
  
  \subsection{SM parameters}
 
A detailed review and discussion of the general issue of finite-$T$ electroweak theory and its ``phase transition" from symmetric to the broken phase is obviously beyond the aims of the present paper. Let us just say that even in the relatively weakly coupled electroweak theory one still finds elements of
 a strongly interacting gauge magnetic sector,  and  other nonperturbative issues, which are reminiscent of QCD phase transitions.
 Here we limit ourselves to the discussion of much simpler effects, such as the lowest-order rescatterings or ``thermal masses", which are dominant, as we are not too close to $T_c$. We will begin with the Higgs boson and then proceed to gauge quanta and finally to fermions. 
 
The temperature dependence  of the effective Higgs potential $V_{eff}(v,T)$ is quite an involved subject, which has been studied by renormalization group methods as well as
numerical lattice simulations.
Let us just say that at high temperatures the Higgs mass and VEV are significantly renormalized, eventually melting away at a critical temperature $T=T_c$. For practical applications, not too close to $T_c$, the Higgs mass and VEV scale down in the same way, approximately as 
\be {
{m_H^2(T)\over m_H^2(0)} =  {v^2(T)\over v^2(0)}= 1-{T^2 \over T_c^2}.  
\label{eqn_v(T)}
}\ee
The masses of W-bosons and quarks receive additional contributions to which we turn now.  The first effects are lowest order rescattering given by the single one-loop diagrams.
 For gauge bosons at  $T$ higher than all masses the result is  \cite{Kapusta}
\be M_W^2 = {1\over 4} g_w^2 v^2(T)+{2\over 3} g_w^2 T^2 + {1\over 6} g_w^2 T^2+ g_w^2 T^2, \ee
where the first term is the usual vacuum mass, but with a $T$-dependent VEV,  and the last
three terms come from $W$ rescatterings on gauge, Higgs and (the single species) quark, by which we mean the top quark. In fact these are due to lowest-order
tree scattering diagrams. Substituting $v(T)$ from (\ref{eqn_v(T)}) one finds that the negative coefficient from a decreasing $v(T)$ and positive coefficients from rescatterings tend to cancel each other! (In fact they do when $T_c\approx 120\, GeV$.) 
Thus we will adopt an  approximation in which $M_W$, unlike $m_H$, does $not$  have a significant dependence on $T$, due to the (diminishing) Higgs part
nearly compensated by the (increasing) ``thermal" part.  We will address how this will affect the binding of $W$'s in the next section.
\\

For  (top) quarks the description of their effective mass is more complicated for several reasons. Similarly to finite-$T$ QCD, the thermal masses due to rescattering on gauge bosons lead to  chirally diagonal $\Sigma_{LL,RR}$ contributions,
 while their interaction with Higgs generates the usual
left-right $LR$ mass term $M$. The resulting Dirac operator in the chiral basis has the structure
\be  D=\left( \begin{array}{cc}  \Sigma_{LL} & M \\ M^+ &   \Sigma_{RR} \end{array} \right) \ee
with the ``masses" with different  chiral structure. In contrast to QCD, weak interactions produce  the $LL$ term only, while strong leads to symmetric $LL+RR$ contribution.
Furthermore, for quasiparticles with nonzero momentum $\vec p$ one has additional splitting in helicity $\sigma=\vec{\sigma} \cdot \vec{p}=\pm 1$ into ``quark" and ``plasmino"
modes as discovered by Klimov and Weldon in 1970's, leading in total to four distinct modes per quark flavor.  
 For details the reader may consult e.g. the paper by Farrar and Shaposhnikov \cite{Farrar:1993hn}, from which we took  specific expressions to be used. At small momenta
 splitting in helicity goes to zero and  there remains  two values of masses, split in chirality, $  \Sigma_{LL,RR} (\omega,\vec p=0)=-\Omega_{LL,RR}^2/\omega$, where
 \ba \Omega^2_{LL}=   {3\pi\alpha_w \over 8} T^2 \left[1+ {M_U^2 \over 3 M_W^2} + {K M_D^2 K^+ \over 3 M_W^2}\right] 
 \nonumber \\ + {2\pi\alpha_s \over 3} T^2,  \hspace{1cm} \ea
 \ba \Omega^2_{RR}=   {3\pi\alpha_w \over 8} T^2 {M_U^2 \over 3 M_W^2}
  + {2\pi\alpha_s \over 3} T^2,  \ea
  while the LR mass term is still given by the Higgs by
\ba M_U=2g_w {M_U \over M_W} \phi \ea
for generic up $U$ quark flavors. If $\phi$ is a function of coordinates, representing the bag, these expressions provide the form of the Dirac eqn for the bound states.

Now, in the absence of gluons,  and specifically for the top quark one finds that the Higgs mass part is much larger than the scattering part
\be M_t^2(T)\approx  M_t^2(0)\left(1-{T^2\over T_c^2}\right)  \gg  \Omega^2_{LL} \sim 0.1 T^2 \ee
which in turn is larger than the  $RR$ component.  Therefore, the coefficient of $T^2$ coming from the Higgs term is factor 20 larger than that coming from rescattering.
If the latter is neglected, one finds a simple approximation for the tops, in which all  $T$-dependence of the top quark comes from the Higgs sector.
If so, selecting e.g. $v(T)$ to define a scale in a binding problem, one finds $M_H(T),M_t(T)$ to (approximately) scale in the same way: which means
that the Dirac equation for the bound levels remains to a good accuracy unchanged, except for the absolute scale. Thus all the results we reported above about 
tops in the bags remains valid, among them the unfortunate conclusions of absence of pure top bags, at any $N$, and thus they must coexist with certain number
of the gauge quanta to form a bag.

Note that in the previous discussion we have considered  the gauge and Yukawa coupling constant to be constant: for completeness, let us comment now 
on what is known about their running.
 The  running of the effective coupling in the 3-d magnetic sector at small momenta $k$. Its renormalization group flow is determined by \cite{Bergerhoff:1996jw}
\begin{equation} 
\frac{\partial g^2(k,T)}{\partial ln(k)}= -{23 \tau \over 24 \pi} g(k,T)^4 {T \over k } + ... 
\end{equation}
which gets strong at momenta smaller than some critical value  $k<k_c$, where the 3d magnetic theory goes into a non-perturbative confining regime. However this is only become important very close to $T_c$ and thus  will not be included.

The next issue is the T-dependence of the Yukawa couplings, especially the largest one defining the top quark mass $g_t$.
As noticed by Marciano \cite{Marciano:1989xd}, its beta function has two terms with an opposite sign: positive from self-coupling and negative, containing $\alpha_s$, from a virtual gluon. They tend to cancel each other, producing a quasi-fixed point.  Furthermore, as a coincidence,the fix point value is not far numerically from the physical value of the coupling for the top quark (but not other quarks). As a result, one can neglect the running of the top Yukawa coupling.

\vspace{2mm}

 Our discussion in this section so far  has been limited to expressions derived in thermal equilibrium. However the specific  application we will discuss in  \cite{FS_baryo}  corresponds to the so called hybrid scenario
  in which there are strong deviations from equilibrium: thus we would like to comment how those can be included. 
  Some of the non-equilibrium effects -- for example the fact that  at the time under consideration gluons are not yet effectively excited -- can be taken care of by simply omitting 
  the corresponding contributions. Some others can be included by treating differently the ``bulk" and  the ``inside"
  of the $WZ-top bags$, by using the same thermal expressions but with different temperatures and densities.
 In the approximation in which the scattering amplitudes can be considered to have a weak enough dependence on the
 momenta, so that they can be factored out of the thermodynamical integrals, one can simply 
use the hard-thermal loop expressions, with $T^2$ substituted by
 by bosonic or fermionic integrals
\be T^2|_{bosonic}  \rightarrow 24 \int {d^3p \over 2 E \, (2\pi)^3} n_p\ee
\be T^2|_{fermionic}  \rightarrow 48 \int {d^3p \over 2 E \,(2\pi)^3} n_p\ee
in which the occupation number $n_p$ is  Bose/Fermi distribution $including$ particle masses  and non-zero chemical potentials.
In particular, one can treat this way the top-gauge scattering inside the bags, in which the density of tops and $W,Z$ quanta are
created occasionally, by large fluctuations in the Higgs field in process of its primordial formation.

\vspace{5mm}

\subsection{$W$ bags at finite $T$}

As we already discussed in the preceding section, at finite $T$ there are two terms in the effective gauge boson mass,
$$ M_W^2 ={g^2\over 4} v^2(T)\phi(r) + M_2(T)^2 $$
where the first term is Higgs-induced and the second term is generated by rescattering on the medium.
The  first term is due to the Higgs coupling and therefore space-dependent via the bag profile function $\phi(r)$. The second term is however
 assumed to be just a position-independent constant, which therefore can be absorbed into the effective frequency
 \be \omega^2 \rightarrow \tilde\omega^2=\omega^2- M_2(T)^2 \ee 
 in all the equations of the preceding section. Furthermore, by using $v(T)$ as the unit of scale, we are left with universal equations which have only $T$-independent profiles, from which universal binding for $\tilde \omega/v(T)$ combinations follow. However, while minimizing the bag, one has to return back to the usual energies.

 We noticed before that in the high-$T$ quadratic approximations, the $O(T^2)$
terms in the first and the second term tend to approximately cancel each other. However the rescaling we are  discussing now works for any $T$-dependence, and  
we do not use this specific approximation. The  expression for the $W$ mass we use (in units of $m^2=(100\, GeV)^2$ as elsewhere) is $m_W^2(T)=.24 T^2+.64$,
so the absolute magnitude of the $W$ mass is growing with $T$.  Regarding the $T$-dependent $decreasing$ Higgs mass, this effect is even more pronounced.

\begin{figure}[h]
\includegraphics[width=2.7in]{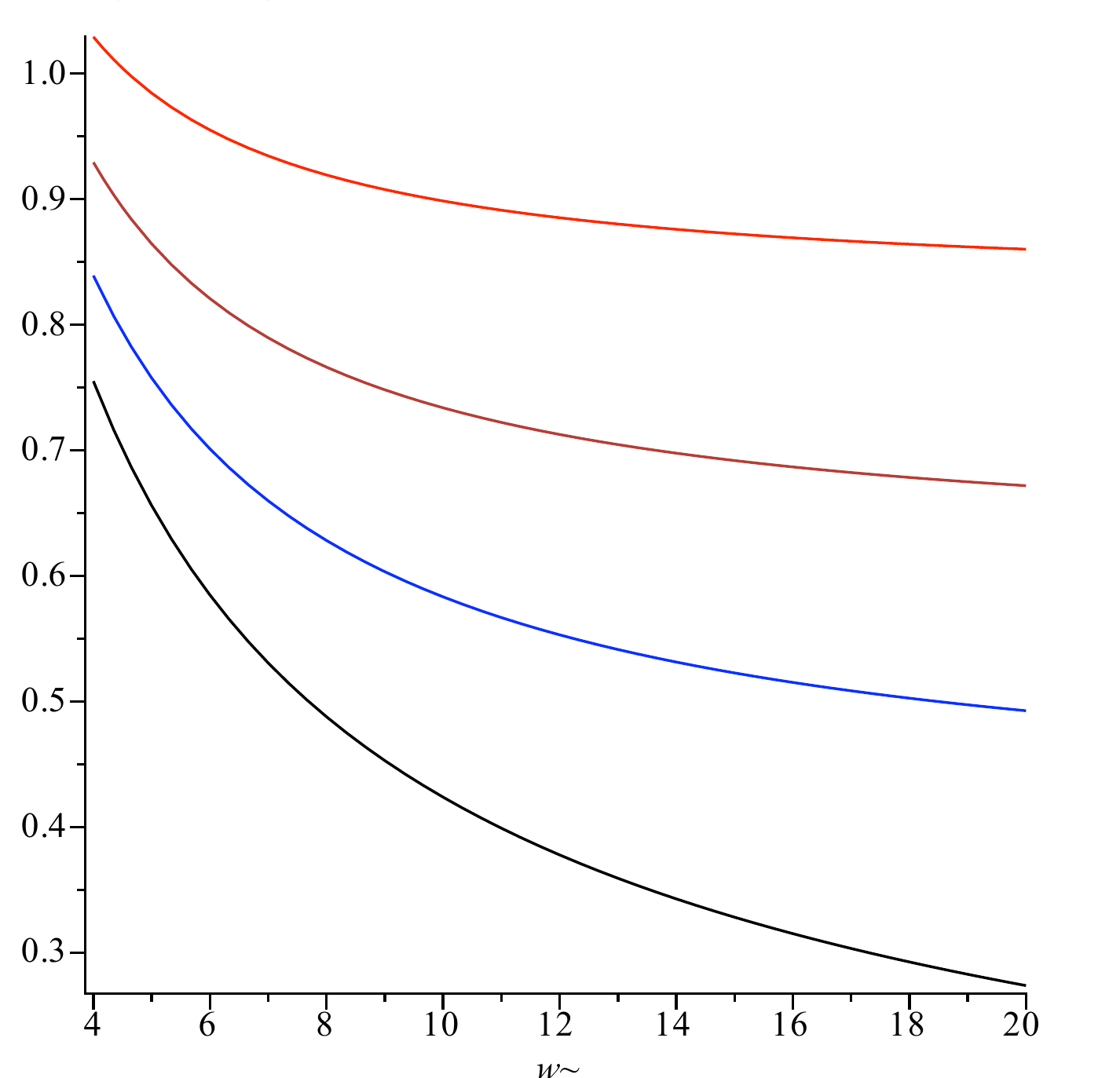} 
\caption{ The lowest $j=1$ magnetic modes in a Gaussian bags versus its size $w$. The four curves, from bottom up,
correspond to $T/(100\, GeV)=0,0.5,0.75,1$, respectively. }
\label{fig_Wbags_atT}
\end{figure}

Using this approximation one can convert the universal spectrum of $\tilde\omega$ for the lowest j=1 magnetic modes in a bag into absolute units. 
Fig.\ref{fig_Wbags_atT} shows a sample of the results. As one can see, the binding energy is reduced as the temperature grows. In our cosmological application in \cite{FS_baryo} we will have $T_{bulk} \sim 50\, GeV$, corresponding to the second curve from the bottom, which is to be compared with the zero $T$ case shown by the bottom curve.   

Substituting these results into the expression for the total energy, together with the Higgs bag energy, one can look for a minimum over all parameters of the
ansatz used, as before. What we found is that  the change of the $W$ mass does not significantly change the pattern of $W$ bag binding. As N increases, W-bags develop first a local minimum for some $N>N_{\ast}$ and become bound for $N>N_c \sim O(1000)$. 

A more important problem for the stability of the bags is of course ``self-ionization", i.e.,  the thermal excitation of a particle from the lowest bound level to the continuum spectrum. For low temperatures, such process will be suppressed by a Boltzmann factor. As $T$ increases, however, the binding energy becomes comparable to the temperature and self-ionization  is no longer suppressed, becoming quite substantial. Indeed, for  $T\sim 50\, GeV$ (the second curve from the bottom in Fig.\ref{fig_Wbags_atT}) one finds that the binding energy per $W$ is $\sim 30\, GeV$. Therefore, the prospect of having long-lived $W-Z-t$ bags at higher $T$  is slim. This is why we focus on ``hybrid" or ``cold" scenario of baryogenesis in  \cite{FS_baryo}.

\section{Summary and Discussion}

We have studied the existence of multi-quanta bags in the electroweak sector of the SM. We found that pure W-bags exist provided the number of gauge bosons is greater than $N_{c}\sim O(1000)$, with typical sizes of the order of $R \sim 0.1 GeV^{-1}$ .  They can be either of  the no-Higgs type, with W-bosons occupying a volume where the Higgs VEV has been depleted from, or of the inverted-type, with W-bosons localized in a surface shell of width $O(1/m_H)$ where the Higgs vanishes, enclosing an inverted-sign Higgs vacuum. Although pure top bags  are excluded from the vacuum, top quarks will certainly bind to these W bags and we found their energy levels in such bags.  Envisioning cosmological applications in our companion paper \cite{FS_baryo}, we have also studied such objects in a finite temperature environment. Although the main properties of such bags are  unchanged at finite temperature, as temperature grows the ionization of these objects becomes important, eventually evaporating them into disappearance at temperatures close the electroweak scale.


\vskip .25cm \textbf{Acknowledgments.} \vskip .2cm The work of PMC and ES
was supported in parts by the US-DOE grant DE-FG-88ER40388.  The work of VF
was supported by the Australian Research Council and NZ Masden fund. ES thanks H.~B. Nielsen for inspiring talk, which made us all to think
about multi-top systems.

\appendix

    \section{$W$-bosons in a Higgs Background} 
    \label{Appendix_bosons}
    
    \subsection{3+1 dimensions}
    
    \subsubsection{Lagrangian, Conserved Charge and Equations of motion}
    Consider boson fields in the electroweak sector of
    the Standard Model (see, e.g., Ref.\cite{weinberg_2001}) 
       \begin{eqnarray}
      \label{gauge}
      {\mathcal L}= -\frac{1}{4}\,
      \left(\partial_\mu \boldsymbol{A}_\nu-\partial_\nu \boldsymbol{A}_\mu  +
        g \,\boldsymbol{A}_\mu \times \boldsymbol{A}_\nu\right)^2,
      \\ \nonumber
            -\frac{1}{4}\,
      \left(\partial_\mu { B}_\nu-\partial_\nu {B}_\mu  \right)^2+
      \frac{1}{2}\,D_\mu\Phi^+ D^\mu \Phi~.
    \end{eqnarray}
    Here $\boldsymbol{A}_\mu$ and $B_\mu$ are the triplet of $SU(2)$
    and the $U(1)$ gauge potentials respectively (abridged notation is
    used here).  The covariant derivative $D_\mu\Phi$ takes into
    account that the Higgs field $\Phi$ has a hypercharge $Y=2$, which
    describes its interaction with the $U(1)$ field, and
    transforms as a doublet under $SU(2)$.
    Taking the unitary gauge one can present it via one real component
    \begin{eqnarray}
      \label{vacuum}
      \Phi =v \left( \begin{array} {c} 0 \\ \phi
      \end{array} \right)~,\quad \phi=\phi^+~.  
    \end{eqnarray}
    Assuming that the scalar field develops the vacuum expectation
    value $\phi=1$ and the Higgs mechanism takes place, one finds
    that the gauge field can be presented as a new $U(1)$ field
    $A_\mu$, and a triplet of massive fields $W^\pm_\mu, \,Z_\mu$
    \begin{eqnarray}
      \label{Amu}
      A_\mu &=&-\sin \theta \,A_\mu^3+\cos\theta \,B_\mu~, 
      \\ \label{Zmu}
      Z_\mu &=& ~~\cos \theta \,A_\mu^3+\sin\theta \,B_\mu~,
      \\ \label{Wmu}
      W_\mu &=&~~\left(A_\mu^1-iA_\mu^2 \right)/\sqrt 2~.
    \end{eqnarray}
    Here $W_\mu\equiv W_\mu^-$ represents the $W$-boson with electric charge
    $e=-|e|$, and $W_{\mu}^{\ast}=W_{\mu}^{+}$  the $W$-boson with charge
    $e=+|e|$ and $\theta$ is the Weinberg angle.  Expanding the Lagrangian Eq.(\ref{gauge}) in the vicinity of
    $Z=A=0$ and retaining only bilinears in the fields
    $W_\mu,W_\mu^+$ terms,
 one derives an effective Lagrangian
    \begin{eqnarray}
      \nonumber
      {\mathcal L}^W = -\frac{1}{2}\left(\partial_\mu W_\nu -
        \partial_\nu W_\mu\right)^{\ast} \left(\partial^\mu W^\nu-
        \partial^\nu W^\mu\right)  
      \\   \nonumber 
      \label{W}      
        +M_W^2\phi^2 \,W_\mu^\ast W^\mu~,
    \end{eqnarray}
    which describes the propagation of $W$-bosons in an external
    Higgs field. As mentioned in the text, due to the global symmetry $W_{\nu} \rightarrow e^{i \alpha} W_{\nu}$ one finds the conserved current 
    \begin{IEEEeqnarray*}{rCL}
j^{\mu }&=&i\Big[ W_{\nu }\left( \partial ^{\mu }W^{\nu \ast }-\partial ^{\nu
}W^{\mu \ast }\right) \\ \nonumber 
&&-W_{\nu }^{\ast }\left( \partial ^{\mu }W^{\nu
}-\partial ^{\nu }W^{\mu }\right)\Big] 
\end{IEEEeqnarray*}
Assuming that the fields are stationary with frequency $\omega$, i.e.
\begin{equation}
W_{\nu}(x,t)=e^{-i \omega t} W_{\nu}(x)
\end{equation}
one finds explicitly  the charge density 
\begin{eqnarray}
j^0&=&- 2 \omega |W_{i }|^{2}-i \left( W^{i}\partial
_{i}W_{0}^{\ast }-W_{i}^{\ast }\partial ^{i}W_{0}\right), 
\label{current density}
\end{eqnarray}

    From the Lagrangian one derives the classical equation of
    motion for gauge bosons
    \begin{eqnarray}
      \left( \Box+M_W^2\phi^2\right) W^\mu 
     - \partial^\mu \partial^\nu \,W_\nu =0~.
     \label{WEOM}
    \end{eqnarray}
    Taking a covariant derivative in Eq.(\ref{wave}) one finds
    \begin{eqnarray}
      \label{Lgauge}
      \partial_\mu (\phi^2 W^\mu)= \phi^2 \partial_\mu  W^\mu
+ W^\mu \partial_\mu (\phi^2)
=0~.
    \end{eqnarray}
  Evaluating $\partial_\mu W^\mu$ from
    Eq.(\ref{Lgauge}) and substituting the result back into
    Eq.(\ref{wave}) one rewrites the latter one in a more transparent
    form
    \begin{eqnarray}
      \left( \Box+M_W^2\phi^2\right) W^\mu
     +\partial^\mu \left(\frac{ W^\nu\partial_\nu \phi^2}{\phi^2}\right) =0.~~\quad
    \end{eqnarray}
    
    This form of the equation of motion will be more suitable for studying the behavior of different W modes close to zero's of $\phi$.
    
    \subsubsection{Hamiltonian}
    
The Hamiltonian density is given as usual by the Legendre transformation
\begin{equation}
\mathcal{H}_{W}=p_{i}\partial _{0}W^{i}+p_{i}^{\ast }\partial _{0}W^{i\ast }-\mathcal{L}_{W}.
\end{equation}%
where $p_{i},\,p_{i}^{\ast }$ are conjugate momenta and are given by
    \begin{eqnarray}
p_{i} &=&-\partial ^{0}W_{i}^{\ast }+\partial _{i}W^{0\ast } \\
p_{i}^{\ast } &=&-\partial ^{0}W_{i}+\partial _{i}W^{0}.
\end{eqnarray}%
and $p_0=p_0^{\ast}=0$ as expected. Using the equations of motion (\ref{WEOM}) and performing some integrations by parts, one finds
\begin{IEEEeqnarray}{rCL}
\mathcal{H}_{W}&=&-\omega \Big[ 2\omega |W_{i}|^{2}+i ( W^{i}\partial _{i}W_{0}^{\ast }\\ \nonumber
&&-W^{i\ast}\partial _{i}W^{0})\Big].  \\ \nonumber
\end{IEEEeqnarray}
Comparing to (\ref{current density}) one finds $\mathcal{H}_W=\omega j^0$. Finally, integrating over space
\begin{equation}
H_{W}=N \omega,
\end{equation}
 which is the desired relation. 
 In order to evaluate the total energy of the bag we must  find the eigenvalues $\omega$ from the equations of motion for the occupied modes and also add
 the pure Higgs terms in the Hamiltonian (the kinetic energy plus $V(\phi)$ which has not been present in the formulae above.
 
    \subsubsection{Spherical coordinates}
We consider a static, spherically symmetric Higgs field $\phi(r)$. The angular dependence of a vector particle wave function will then be given by the spherical vectors (see, e.g., \cite{LL4})
    \begin{eqnarray}
      \label{Y}
      \begin{array}{l}
      \boldsymbol{ Y}^{(e)}_{jm}= \boldsymbol{
        \partial}_n\,Y_{jm}/\sqrt{j(j+1)}~,
      \\
      \boldsymbol{ Y}^{(l)}_{jm}= \boldsymbol{ n} \,Y_{jm}~,
      \\  
      \boldsymbol{ Y}^{(m)}_{jm}= \boldsymbol{ n \times Y}^{(e)}_{jm} 
        ~.
      \end{array}
    \end{eqnarray}
    Here $Y_{jm}\equiv Y_{jm}(\theta,\varphi)$ is the spherical
    function and, $\boldsymbol{ Y}^{(e)}_{jm},\boldsymbol{ Y}^{(l)}_{jm},
    \boldsymbol{ Y}^{(m)}_{jm}$ are the electric, longitudinal and
    magnetic vectors. The symbol $\boldsymbol{ \partial}_n$ in
    Eq.(\ref{Y}) indicates the angular part of the gradient, $
    \boldsymbol{ \partial}F(\theta,\phi)= \boldsymbol{ \partial}_n
    F(\theta,\phi)/r$, and $\boldsymbol{ n}=\boldsymbol{ r}/r$ is a
    unit vector along the radius vector. Relevant formulas for the Laplace operator read
      \begin{eqnarray}
        \label{ddd}        
     \begin{array}{l}
     \Delta_n\boldsymbol{ Y}^{(\,e\,)}_{jm}  = -j(j+1)\,\boldsymbol{ Y}^{(e)}_{jm}+
          2\sqrt{j(j+1)} \, \,\boldsymbol{ Y}^{(l)}_{jm},
          \\ 
           \Delta_n\boldsymbol{ Y}^{(\,l\,)}_{jm}=2\sqrt{j(j+1)}\,\,
          \boldsymbol{ Y}^{(e)}_{jm}
          - \big(j(j+1)+2\big) \boldsymbol{ Y}^{(l)}_{jm},
          \\  \nonumber
          \Delta_n\boldsymbol{ Y}^{(m)}_{jm} =-j(j+1)\boldsymbol{
          Y}^{(m)}_{jm}.
        \end{array}
    \end{eqnarray}
    Here $\Delta_n$ describes the angular part of the Laplacian, i.e.
    $\Delta F(\theta,\phi) =\Delta_nF/r^2$. The parity for electric
    and longitudinal polarizations equals $P=(-1)^{j}$, for magnetic
    polarization the parity is $P=(-1)^{j+1}$. The orbital moment $l$
    takes the value $l=j$ for the magnetic polarization, in agreement
    with the parity for this state (states $p_1, d_2,f_3 \ldots$).
 
 The electric and longitudinal polarizations are constructed as linear combinations of the two
    states with $l=j\pm 1$: this is why they are mixed by the
    Laplacian operator. For $j=0$ there exists only one spherical
    vector, which is purely longitudinal and has $l=1$.
  The mixed electric-longitudinal modes
    correspond to the following states with $j\ge 1$: $s_1, p_2,
    d_1, d_3, f_2 \ldots$. The states with $j=0$ are given by the
 longitudinal mode alone (states $p_0$).  

   For a static, spherical Higgs field $\phi(r)$, Eq. (\ref{Lgauge}) gives
 \begin{eqnarray}
      \label{LgaugeS}
 \phi \partial_\mu  W^\mu= 2 {\bf W \boldsymbol{ n}} \frac{d \phi}{d r}.
    \end{eqnarray}
The electric $\boldsymbol{ Y}^{(e)}_{jm}$ and magnetic  
    $\boldsymbol{ Y}^{(m)}_{jm}$ vectors are orthogonal to
$\boldsymbol{ n}$ (${\bf W \boldsymbol{ n}}=0$), while for the longitudinal wave ${\bf W \boldsymbol{ n}}=W$.

\subsubsection{Magnetic modes}

\begin{figure}[t]
\includegraphics[width=2.7in]{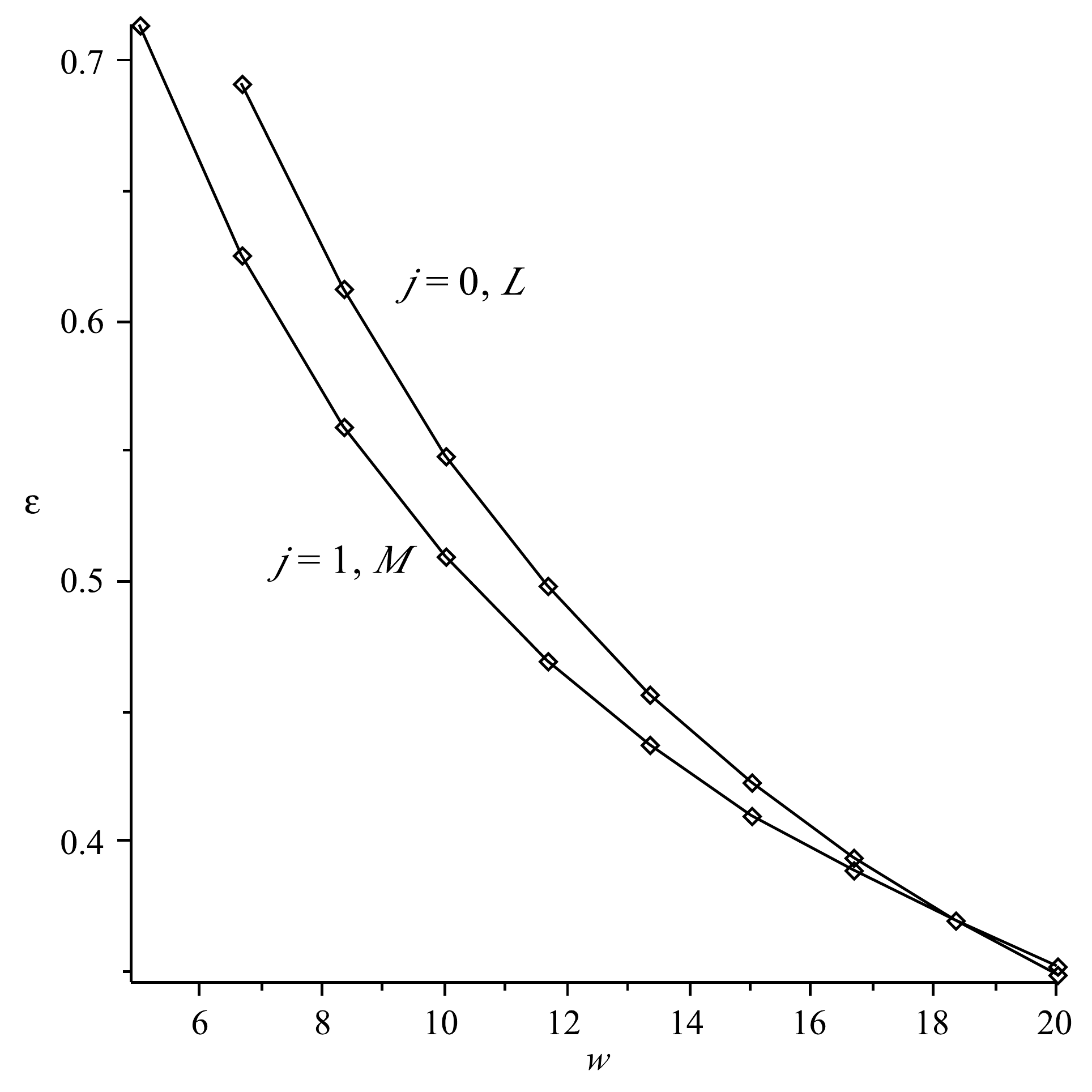}
\caption{The energy of the  levels $\varepsilon/(100 \,GeV)$ versus the size of the bag
$w\cdot (100\, GeV)$, for the Gaussian
ansatz with $\alpha=0.9$ (no zero case). The $ j=1$ M-mode is below the $j=0$ L mode.}
\label{fig_levels}
\end{figure}

For a magnetic polarization,  Eq.  (\ref{LgaugeS}) tells us that  $\partial_\mu  W^\mu=0$. 
Therefore, the last term in  Eq.(\ref{form})  vanishes and $f_m$ satisfies the simple Klein-Gordon equation (\ref{magnetic}) with $j\geq1$. Although also true for the electric polarization, due to the fact that the Laplacian mixes the electric and longitudinal polarizations one finds the set of coupled equations (\ref{electric}, \ref{longitudinal}). As mentioned in the text, this fact makes us expect that the magnetic modes are below the electro-longitudinal ones. In order to verify this,  we numerically solved the W-boson spectrum in the background (\ref{eqn_Gaussian}) for various values of $\alpha$. The results for a particular example with $\alpha=0.9$ are shown in Fig. \ref {fig_levels}. Comparing the $j=1$ magnetic mode (large circles) with the $j=0$ longitudinal mode (large boxes) one can see that the former levels are lower, in spite of having a larger $j$. This is the effect of the repulsive potential, which moves the longitudinal states upward, even when no zero and singularities are present. Therefore, we fill only the lowest  ($j=1$) magnetic level when searching for W- bag solutions. 

\subsubsection{Electro-longitudinal modes}

  For the longitudinal mode the effective potential $\partial^\mu (\frac{ W^\nu\partial_\nu \phi^2}{\phi^2})$ remains and becomes singular,  $\sim 1/{\phi^2}$, near $\phi=0$.   It is easy to understand the reason for this singular behavior. The longitudinal wave does no exist for a zero mass particle. Therefore, the longitudinal wave must vanish in the area $\phi=0$ since the Higgs field $\phi$ plays a role of the effective mass.

It is convenient to use the substitution
\be   L(r)={\psi(r) \over r \phi(r) } \ee
to recast Eq.  (\ref{longitudinal}) into the following Schreodinger-like form
\begin{multline}
\omega^2 \psi(r)=-{d^2 \psi(r) \over dr^2} +V(r)\psi(r)\\
+2\phi(r)\sqrt{j(j+1)}f_e(r)/r^2) 
\end{multline}
with the last mixing term and the effective potential
\begin{multline}
V(r)={2+ j(j+1)\over r^2} +\phi(r)^2  M_W^2   -{d^2 \phi(r) \over dr^2} {1\over \phi(r)} \\ \nonumber 
 +2{d\phi(r)\over dr}{1 \over r\phi(r)} +  2{(d \phi(r)/dr ))^2 \over \phi(r)^2} 
\end{multline}
It now includes  a new term with the first derivative of the kink: yet the most singular term at the position of a zero is with the square
of the  first derivative of $\phi(r)$. This $1/(r-r_0)^2$ singularity makes the wave function vanish at $r_0$, which keeps the
energy finite.

 Moreover, the corresponding barrier is basically impenetrable for the longitudinal $W$. Indeed, in the semiclassical approximation one can see that the action $\int p(r)dr$ diverges. There are  basically two sets of levels; inside and outside of the bag.

  \subsection{1+1 dimensions}
  
    The content of this section is not strictly speaking relevant to the discussion of gauge quanta binding to a bag. It serves only the purpose of illustrating what exactly happens with the longitudinal mode in the simplest case of a very large bag $\phi(x)$ possessing a zero (approximated by a 1-dimensional kink-like Higgs profile). We will then return to the finite-size bag with a Higgs zero.

 The 1D kink separates two vacua,  with  opposite signs of the Higgs VEV. By symmetry, $\phi(x)$ is assumed to be an odd function, being zero at the location of the kink, which we take at $x=0$. Therefore, the effective mass of SM particles is zero near $x=0$, therefore creating attraction to this point. Let us discuss gauge bosons moving only in $x$ direction normal to the kink. Transverse polarization modes $E,M$ are decoupled from the longitudinal one and satisfy the same equation of motion, namely,
\ba 
-{d^2 F(x) \over dx^2}  +  M_W^2\phi(x)^2 F(x)=\omega^2 F(x).
\ea
 The effective potential is thus $M_W^2\phi(r)^2$, just as it would be for a scalar field. If $\phi(x)\sim x$ at small $x$, this potential is quadratic, with oscillatory-like levels with positive squared frequency $\omega^2$.
 
The longitudinal polarization, however, contains several additional terms 

\begin{multline} 
-{d^2 L(x) \over dx^2}  +  M_W^2\phi(x)^2 L(x) 
-{2\over \phi(x)} {dL \over dx}  {d\phi \over dx} \\ \nonumber
 - {2 \over \phi} {d^2 \phi \over dx^2} L(x) + {2\over \phi^2} {d\phi \over dx} L(x)=\omega^2 L(x).
\end{multline}
The first derivative of the function may be removed by the following substitution 
\be 
L(x)={\psi(x) \over \phi(x)},
\ee
leading to the Schrodinger-like equation for the new wave function $\psi(x)$
\ba
 \omega^2 \psi(x)=- {d^2 \psi(x) \over dx^2} 
+M_W^2\phi(x)^2) \\ \nonumber
-{d^2\phi \over dx^2} { \psi(x)\over \phi(x)}  +2 \psi(x) {(d\phi(x)/ dr)^2 \over \phi(x)^2}  
\ea
 Note that the new terms in the effective potential are singular at zeros of $\phi(x)$.
The term with the second derivative of the kink shape $\phi(x)$ is
 harmless because a generic expansion of an odd function
$\phi(x)=C_1 x+C_3 x^3+...$ with some finite constants $C_1,C_3$ will only produce a constant term in the effective potential, equal
to $6C_3/C_1$. Yet the other term has a very significant singularity.

\begin{figure}[t]
\includegraphics[width=2.7in]{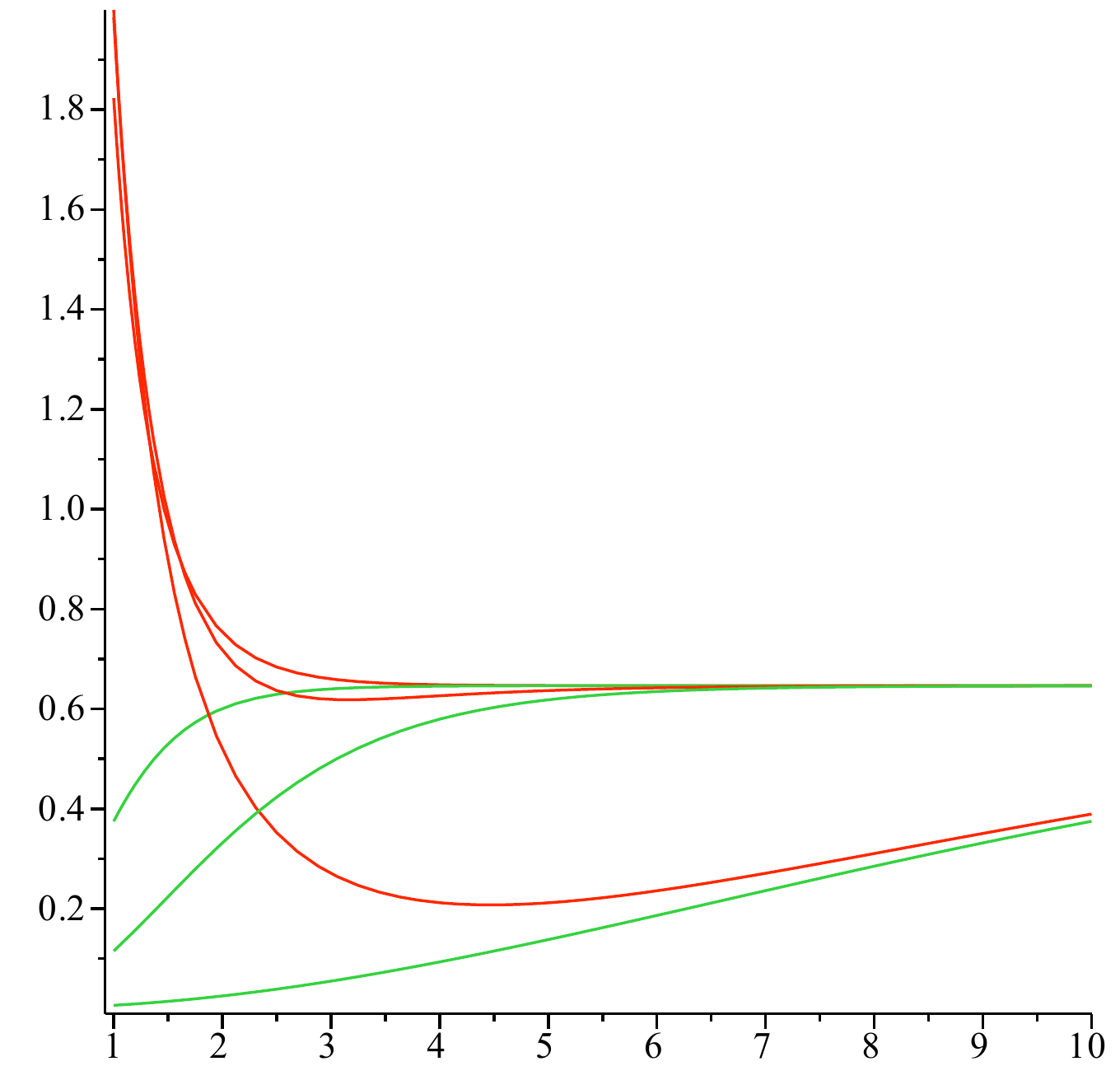} 
\caption{The effective potential for the  $W$ levels $V/(100 \,GeV)$ versus the distance
$x(100\, GeV)$ from the kink center. The lower (green) curves are for the M modes, the upper (red) ones are for the longitudinal L ones. 
Three cases shown  are for $A/(100\, GeV)=0.1,0.45$ and $1$, from bottom to top, respectively.  }
\label{fig_pots}
\end{figure}

Now, let us take for definiteness a particular kink profile, e.g., 
\be \phi(x)= tanh(Ax)\ee
and see what modification the new term introduces to the effective potential. The effect depends crucially on the 
kink's width, or parameter $A$. This is demonstrated in Fig. \ref{fig_pots}: as one can see in the lowest pair of curves, for $A/(100\, GeV)=0.1$,
case there remains some attractive part of the kink, while in   the $A/(100\, GeV)=1$ Upper) case there remains only repulsive  potential,
with the middle case of  $A/(100\, GeV)=0.45$ being the critical one. Apparently, the longitudinal mode is well above the
transverse modes, if it even has any bound states at all.

The reader can also be reminded about a comparison with the well known situation with fermions in a kink field, see recent discussion and original refs in \cite{KFS}.
In this case the effective potential also develops a negative part, induced by the spin terms.  However
in this case the lowest level has {\em exactly zero} value. These correspond to the well-known fermionic zero modes, enforced by topological index theorems.

Summarizing the lessons  from the  one dimensional kink problem:   The lowest fermionic levels are in this case at zero energy, lower 
than the zero-point energy for transverse  W-boson  levels. The longitudinal W-bosons are strongly repelled from the zeros of the Higgs and thus their energy levels are much higher than those of transverse modes.

\section{Fermions in a Higgs Background}
\label{Appendix_fermions}
\subsection{\protect\bigskip 3+1 dimensions}

\subsubsection{Spherical coordinates}
We take signature $(+---)$ and use the standard representation for Dirac matrices 
\begin{equation}
\gamma ^{0}=\left( 
\begin{array}{cc}
1 & 0 \\ 
0 & -1%
\end{array}%
\right) ,\quad \gamma ^{i}=\left( 
\begin{array}{cc}
0 & \sigma ^{i} \\ 
-\sigma ^{i} & 0%
\end{array}%
\right)
\end{equation}%
and $\bar{\psi}=\psi ^{\dagger }\gamma ^{0}$.   In two component notation
\begin{equation}
\psi=\left( 
\begin{array}{c}
\eta  \\ 
\chi 
\end{array}%
\right),
\end{equation}
In spherical coordinates, we write%
\begin{equation}
\left( 
\begin{array}{c}
\eta  \\ 
\chi 
\end{array}%
\right) =\frac{1}{r}\left( 
\begin{array}{c}
F(r)\Omega _{jlm} \\ 
(-1)^{1/2(1+l-l^{\prime })}G(r)\Omega _{jl^{\prime }m}%
\end{array}%
\right) ,
\end{equation}%
with normalization $\int dr \,(F^{2}+G^{2})=1$ and where  $\Omega _{jlm}$ are spherical 2-component spinors satisfying%
\begin{equation}
\Omega _{jl^{\prime }m}=i^{1-l^{^{\prime }}}(\vec{\sigma}\cdot \hat{r}%
)\Omega _{jlm}
\end{equation}%
and 
\begin{equation}
\int do\Omega _{ljm}^{\ast }\Omega _{l^{\prime }j^{\prime }m^{\prime
}}=\delta _{ll^{\prime }}\delta _{jj^{\prime }}\delta _{mm^{\prime }}.
\end{equation}%
\subsubsection{Hamiltonian}
The Dirac Hamiltonian is given by
\begin{eqnarray}
H_D=\int d^3 r \, \psi^{\dagger} \,h_D \, \psi, 
\end{eqnarray}
with
\begin{eqnarray}
h_{D}=\gamma ^{0}\left( -i\gamma ^{i}\partial _{i}+m\phi \right).
\end{eqnarray}
Therefore, 
\begin{multline}
H_{D}=\int d^{3}r\Big[\left( -i\eta ^{\ast }(\vec{\sigma}\cdot \vec \partial)\chi
+\text{h.c.}\right)  \\
+m\phi \left( \eta ^{\ast }\eta -\chi ^{\ast }\chi \right)\Big].
\end{multline}
Using the spherical spinor properties one can show that 
\begin{equation}
(\vec{\sigma}\cdot \vec{p})\chi =-\frac{1}{r}\,\left(G^{\prime }-\frac{\kappa }{%
r}G\right),  \label{property spinors}
\end{equation}%
(and similarly for $\eta$) which leads to%
\begin{multline}
H_{D}=\int dr\Big[ \left( -G^{\prime }+\frac{\kappa }{r}G\right) F+\left(
F^{\prime }+\frac{\kappa }{r}F\right) G  \\
+m\phi \left( F^{2}-G^{2}\right)\Big] .
\label{Dirac Hamiltonian 1}
\end{multline}
If $\psi _{a}$ denotes a spinor solution with \textit{positive} energy $%
\varepsilon _{a}$, i.e. $h_{D} \, \psi _{a}=\varepsilon _{a}\psi _{a}$ then, using (\ref{property spinors}), the equations of motion for $F$ and $G$ read
\begin{eqnarray}
\left( \varepsilon _{a}-m\phi \right) F_{a} &=&-G_{a}^{\prime }+\frac{\kappa 
}{r}G_{a}, \label{EOMF} \\
\left( \varepsilon _{a}+m\phi \right) G_{a} &=&F_{a}^{\prime }+\frac{\kappa }{%
r}F_{a}.
\label{EOMG}
\end{eqnarray}%
It is readily seen from  (\ref{Dirac Hamiltonian 1}) that using Eqs. (\ref{EOMF}, \ref{EOMG}), the total fermionic energy is again given simply by
\begin{equation}
E_{\psi}=\sum_{a}n_{a}\varepsilon _{a}
\end{equation}%
where $n_{a}$ is the occupation number of each state and $N=\sum_{a}n_{a}$
is the total fermion number. Due to charge conjugation symmetry, we shall
consider $n_{a}>0$ and $\varepsilon _{a}>0$ for both quarks and anti-quarks.

\subsection{1+1 dimensions}

We take signature $(+-)$ and use the standard representation%
\begin{equation}
\gamma ^{0}=\sigma^{3}, \quad \gamma ^{1}=\sigma^{2}, \quad \gamma^{2}=\sigma^{1}.
\end{equation}%
for Dirac matrices in 1+1 and $\bar{\psi}=\psi ^{\dagger }\sigma ^{3}$. We write
a Dirac spinor as%
\begin{equation}
\psi =\left( 
\begin{array}{c}
F \\ 
G%
\end{array}%
\right) 
\end{equation}%
with left and right moving spinors
\begin{eqnarray}
\psi _{L}=\frac{1}{2}\left( F-G\right) \left( 
\begin{array}{c}
1 \\ 
-1%
\end{array}%
\right) , \nonumber \\
\psi _{R}= \frac{1}{2}\left( F+G \right) \left( 
\begin{array}{c}
1 \\ 
1%
\end{array}%
\right). \,\,\,\,\,\nonumber
\end{eqnarray}

The Hamiltonian reads%
\begin{equation}
H_{D}=\int dx\left[ F^{\prime }G-FG^{\prime }+m\phi (F^{2}-G^{2})\right]. 
\end{equation}%
Note that the 3d system becomes one dimensional in the $r\rightarrow \infty $
limit, with the indices $(jlm)$ becoming "internal" and $F$ and $G$
becoming the upper and lower components of a 1+1 Dirac spinor. 

\subsection{Numerical results}

In order to see what actually happens with the fermionic levels as
the bag's node moves from large distances towards the origin, we have
numerically solved the Dirac equation in the background of the Higgs ansatz (\ref{kinksolution}).
\begin{figure}[tbp]
\centering
\includegraphics[width=2.7in]{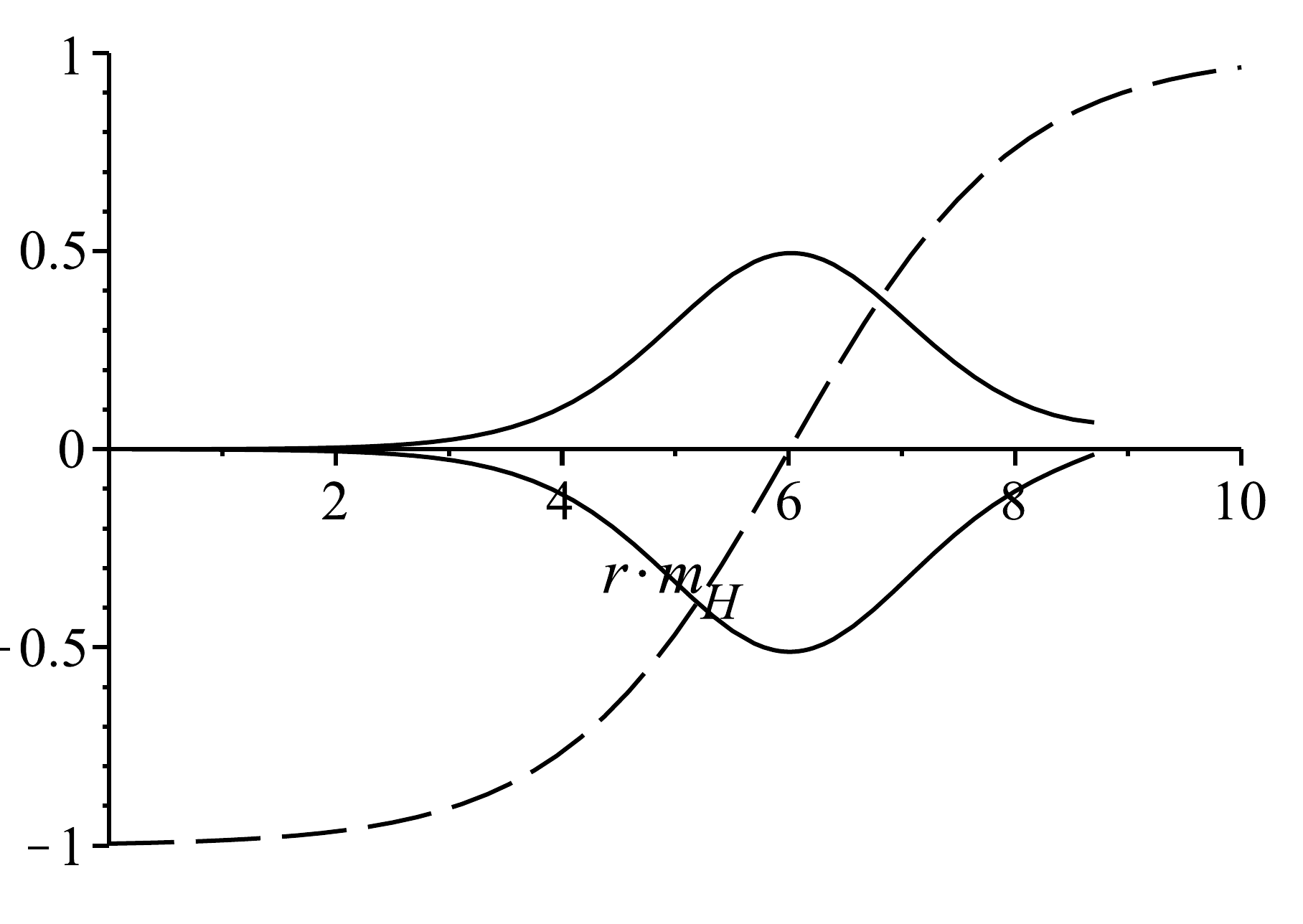}  \includegraphics[width=2.7in]{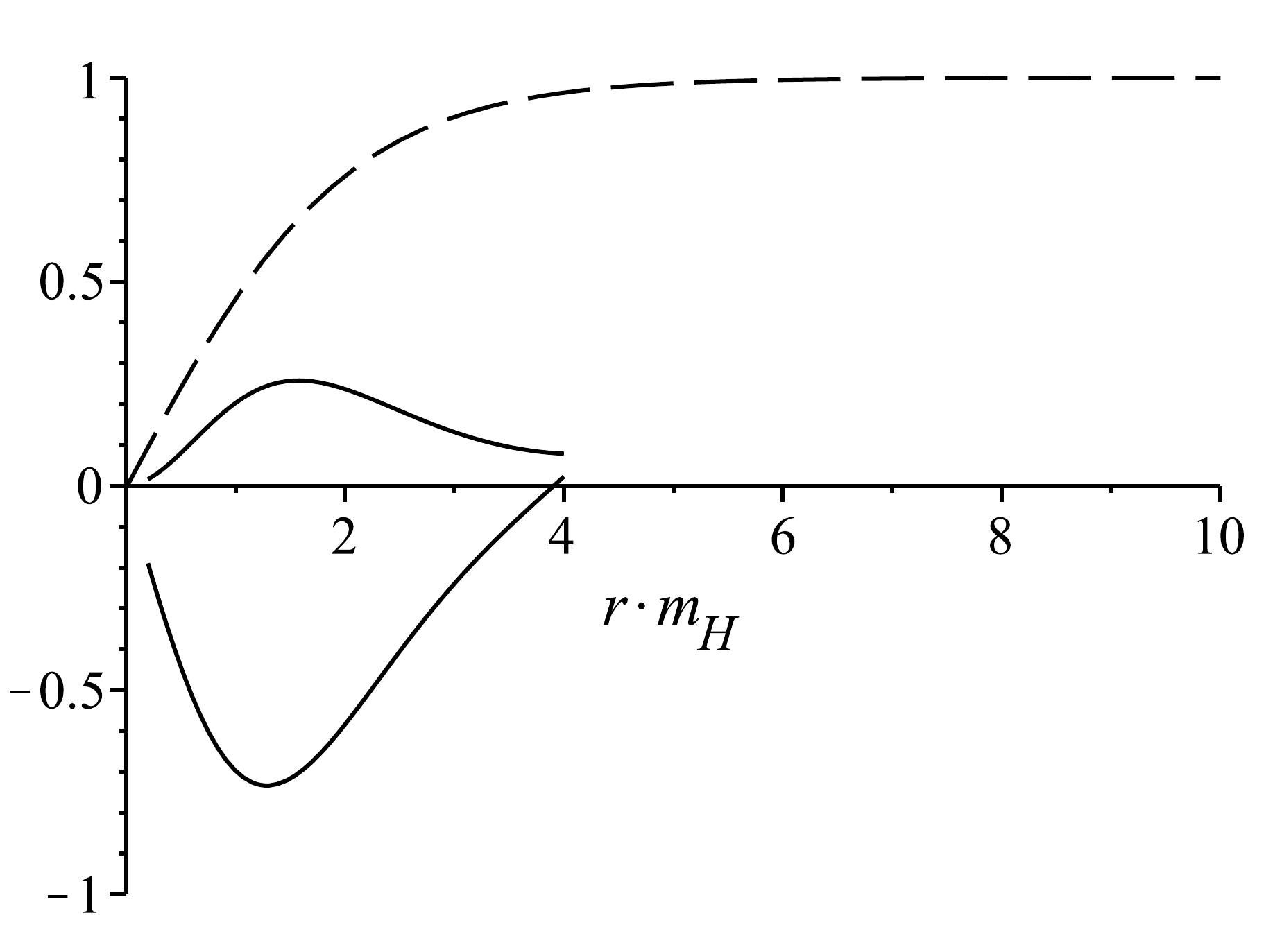} 
\caption{Upper and lower components (upper and lower solid lines) of Dirac's equation in the Higgs background (dashed line). The upper figure shows a large kink profile with the lowest level, $\kappa=-1$, asymptotically becoming a left-moving (i.e. $F=-G$) zero mode. As the node moves towards the origin, the system becomes increasingly non-relativistic and $|F|>|G|$, as expected. A corresponding behavior for an anti-kink profile also occurs, now with the $\kappa=1$ level asymptotically becoming a right-moving (i.e. $F=G$) zero mode.}
\label{fig_kink}
\end{figure}

In Fig.\ref{fig_kink} we follow the $l=0$ Dirac wave functions for a kink profile. When the node is 
located at a rather large distance from the origin, the solution is very close to the analytic Gaussian solution discussed above: the maxima of both $F(r)$ and $G(r)$ are very close to the position of the
Higgs' node and $|F| \sim |G|$. As the Higgs' node moves closer to the origin, although $F(r)$ and $G(r)$ are still approximately Gaussian, their maxima no longer trace the position of the zero and they are visibly different, $|F|>|G|$. This tendency continues as the node moves past the origin and finally disappears. This  
is also easy to understand: as the bag potential becomes more shallow 
the system becomes increasingly non-relativistic, in which case one would indeed expect $|F|\gg |G|$. Apart from the $l=0$ level, it is also interesting to follow the behavior of the energy levels with $l>0$ as a function of $R$.

\cleardoublepage

\end{document}